%% file: h4366.tex
\documentclass{aa}
\usepackage{graphicx}
\usepackage{rotating}

\def\spose#1{\hbox to 0pt{#1\hss}}                                            
\def\simlt{\mathrel{\spose{\lower 3pt\hbox{$\mathchar"218$}}
     \raise 2.0pt\hbox{$\mathchar"13C$}}}
\def\simgt{\mathrel{\spose{\lower 3pt\hbox{$\mathchar"218$}}
     \raise 2.0pt\hbox{$\mathchar"13E$}}}                                     
\def\cgs{\rm erg cm$^{-2}$ s$^{-1}$}
\def\pn{\par\noindent}

\begin{document}

\title{The HELLAS2XMM survey: III.}
\subtitle{Multiwavelength observations of
hard X--ray selected sources in the PKS~0312--77 field\thanks{This work is based
on observations collected at the European Southern Observatory, La
Silla, Chile (proposals ID: 66.A-0520(A), 67.A-0401(A), 68.A-0514(A)) and Paranal, Chile
(proposal ID: 69.A-0506(A)), and observations made with the XMM--{\it Newton}, 
an ESA science mission with instruments and contributions directly funded by ESA member
states and the USA (NASA)}}

\author{M. Brusa \inst{1,2} \and A. Comastri \inst{2} \and 
        M. Mignoli \inst{2} \and F. Fiore \inst{3} \and 
        P. Ciliegi \inst{2} \and C. Vignali \inst{2,4} \and 
        P. Severgnini \inst{5} \and 
	F. Cocchia \inst{3} \and
	F. La Franca \inst{6} \and G. Matt \inst{6} \and
	G.C. Perola \inst{6} \and R. Maiolino \inst{7} \and A. Baldi
        \inst{8} \and S. Molendi \inst{8}}

\offprints{M. Brusa}

\institute{Dipartimento di Astronomia Universit\`a di Bologna, 
    via Ranzani 1, I--40127 Bologna, Italy \\
    \email{brusa@bo.astro.it}
    \and INAF -- Osservatorio Astronomico di Bologna,  
    via Ranzani 1, I--40127 Bologna, Italy \\
    \email{comastri,mignoli,ciliegi@bo.astro.it}	
     \and INAF -- Osservatorio Astronomico di Roma, 
    via Frascati 33, I--00040 Monteporzio, Italy\\
    \email{fiore,cocchia@quasar.mporzio.astro.it}
    \and Department of Astronomy and Astrophysics, 
     The Pennsylvania State University, 525 Davey Lab, 
     University Park, PA 16802, USA \\
     \email{chris@astro.psu.edu}
     \and INAF -- Osservatorio Astronomico di Brera,  
    via Brera 28, I--20121 Milano, Italy\\
     \email{paola@brera.mi.astro.it}
    \and Dipartimento di Fisica Universit\`a di Roma Tre,
    via della Vasca Navale 84, I--00146 Roma, Italy \\
    \email{lafranca,matt,perola@fis.uniroma3.it}  
    \and INAF -- Osservatorio Astrofisico di Arcetri,
     Largo E. Fermi 5, I--50125, Firenze, Italy \\
    \email{maiolino@arcetri.astro.it}
    \and IASF -- CNR, Istituto di Fisica Cosmica,
     via Bassini 15, I--20133, Milano, Italy \\
    \email{baldi,molendi@mi.iasf.cnr.it}}



\abstract{
We present extensive optical, radio and infrared follow-up
observations of a sample of 35 hard X--ray (2--10 keV) selected sources 
discovered serendipitously in the PV XMM--{\it Newton} observation of
the radio--loud quasar PKS~0312--77 field, for which also an archival
{\it Chandra} observation is available. 
The observations have been carried out as part of the {\tt
HELLAS2XMM} survey, a program aimed to understand the nature  
of the sources responsible for the bulk of the hard X--ray Background (XRB).
The identification of the optical counterparts greatly benefits from the 
positional accuracy obtained from {\it Chandra} and radio
observations. As a consequence, the spectroscopic completeness of
the present sample (80\%) is limited only by the faintness
of the optical counterparts.
The multiwavelength coverage of our survey allows us to unveil
a large spread in the overall properties of hard X--ray
selected sources. At low redshift ($z<1$), the source breakdown 
includes Broad Line AGN, Narrow Emission--Line Galaxies, and 
optically ``normal'' galaxies. 
All the ten sources at $z>1$ are spectroscopically
classified as Broad Line AGNs. A few of them show significant intrinsic 
X--ray absorption (N$_H>10^{22}$ cm$^{-2}$), 
further supporting previous evidence of a decoupling
between optical and X--ray properties at high luminosities and redshifts.
Finally, a non negligible fraction ($\sim$15\%) of the hard X--ray
sources are not detected down to the limiting magnitude of the optical images.
The corresponding high X--ray to optical flux ratio, X--ray and
optical--infrared colors strongly suggest that they are high redshift,
obscured AGN.  
}
\maketitle
\section{Introduction}
\par\noindent
Deep X--ray surveys carried out with \textit{Chandra} and
XMM--\textit{Newton} have resolved 
a large fraction (more than 80\%) of the hard X--ray
Background (XRB) into discrete sources 
(Mushotzky et al.~2000; Brandt et al. 2001; Alexander et al. 2003; 
Hasinger et al. 2001; Giacconi et al. 2001) down to a 2--10 keV
flux limit of about $2\times 10^{-16}$ erg cm$^{-2}$ s$^{-1}$.\\
The results of the optical identifications show that about half of 
the objects are associated with optically bright (I$<$24) galaxies at
redshifts  
$<$ 1.5 which are often identified with Active Galactic Nuclei (AGN), 
while the other half appears to be a mixture of higher redshift 
AGN and optically faint (I $>$ 24) galaxies 
(Alexander et al. 2001; Barger et
al. 2001, 2002; Giacconi et al. 2002).
\par\noindent
In order to better understand the nature of the various components of the
X--ray background light we have started a program of multiwavelength 
follow--up observations of hard X--ray selected sources 
serendipitously discovered in 15 XMM--{\it Newton} fields 
over $\sim 3$ deg$^{2}$ 
(the {\tt HELLAS2XMM} survey; Baldi et al. 2002).
The 2--10 keV sample of the {\tt HELLAS2XMM} survey consists of 
495 sources detected in the hard X--ray band at fluxes of the order 
of $\sim 10^{-14}-10^{-12}$ erg cm$^{-2}$ s$^{-1}$ where a significant fraction
of the XRB is resolved ($\sim 50-60$\%; see, e.g., Comastri 2001
for a review) and at the same time the optical identification process 
is relatively easy.
This strategy allows to cover a large area of the sky 
and to fill the gap between previous shallow hard X--ray surveys
(the {\it Beppo}SAX {\tt HELLAS} survey, Fiore et al. 2001;
the ASCA {\tt GIS} survey, Cagnoni et al. 1998;
the ASCA {\tt LSS} and {\tt MSS} Surveys, Akiyama et al. 2000;
Ueda et al. 1999; Akiyama et al. 2002) and recent deep {\it Chandra}
(CDF--N, Brandt et al. 2001; CDF--S, Giacconi et al. 2002) and 
XMM--{\it Newton} (Lockman Hole, Hasinger et al. 2001) observations.\\
To date, we have performed an extensive optical follow--up program for
about one third of the fields (Fiore et al. 2003).
The final aim of this project is the derivation of an accurate
luminosity function over a wide range of redshifts and luminosities
for a large sample of hard X--ray selected, presumably obscured
objects, in order to trace the accretion history in the
Universe. 
Indeed, X--ray absorbed sources 
are a key parameter for AGN synthesis models 
of the XRB, which, in their simplest versions (Setti \& Woltjer~1989; 
Madau et al.~1994; Comastri et al.~1995), predict a
large number of high--luminosity, highly obscured quasars (the
so--called QSO2). 
In the zero--th order unifications models, QSO2 are predicted to be
the high--luminosity counterparts of local Seyfert 2 galaxies.
Despite intensive optical searches, these narrow--line
high--redshift objects appear to be elusive, suggesting a 
space density and evolution different from that expected from unified
schemes and calling for substantial revision of the XRB baseline models 
(Gilli et al.~2001). 
However, the results of recent multi--wavelength follow--up from both 
shallow and deep surveys, indicate that the sources responsible for a large
fraction of the XRB energy density are characterized by a large spread
in their optical properties; therefore the understanding of the 
energetically dominant component of the XRB is possible only by 
means of multiwavelength observations. 
\par\noindent
In this framework, the field surrounding the radio--loud quasar 
PKS~0312--77 (one of the {\tt HELLAS2XMM} fields)
is a key example. 
It has been observed both by {\it Chandra} 
and XMM--{\it Newton} during their Calibration and Performance
Verification (PV) phases;
deep radio observations at 5 GHz have been obtained 
with the Australian Telescope Compact Array (ATCA) telescope, along with
the optical imaging at the ESO 3.6m telescope for the 35 objects 
detected in the combined {\it MOS1} + {\it MOS2} + {\it pn} XMM--{\it
Newton} observation. Optical spectroscopy has been obtained 
for 28 sources, both at the ESO 3.6m and VLT/FORS1 telescopes. \\
In Section 2 we present the multiwavelength data, in Section 3
the X--ray sources identification, and in Section 4 the
radio properties. In Section 5 we discuss the source breakdown and
the multiwavelength properties of the sources.
Finally, in Section 6 we summarize our results.\\
Throughout the paper, the adopted values for the Hubble constant and the
cosmological parameters are H$_0$=70 km s$^{-1}$ Mpc$^{-1}$, 
$\Omega_{\Lambda}$=0.7, $\Omega_{\rm m}$=0.3. 
\section{Multiwavelength observations}
\subsection{XMM--{\it Newton}}
\par\noindent
The PKS~0312--77 field was observed during the XMM--{\it Newton}
PV phase, in 2000, March 31, for a
nominal exposure time of $\sim$ 30 ks.\\
The XMM--{\it Newton} data were processed using version 5.3 of the
Science Analysis System (SAS). The event files were cleaned up from
hot pixels and soft proton flares (see Baldi et al. 2002 for details).
 The resulting exposure times are 24.7, 26.5 and  26.1 ks in the
{\it pn}, MOS1 and MOS2 detectors, respectively.\\
The excellent relative astrometry between the three cameras (within
1$''$, below their FWHM of $\sim6''$) allows us to merge the
\textit{MOS} and \textit{pn} images in order to increase the signal
to noise ratio and to reach fainter X--ray fluxes.\\
An accurate detection algorithm developed by our group (Baldi et al. 2002)
was run on the 2--10 keV cleaned event, in order
to create a list of candidate sources.
We then computed the probability that the detected counts originate from
poissonian background fluctuations:
35 sources were detected above a detection threshold of
{\it p}=2$\times 10^{-5}$.
The count rate to flux conversion factor was derived assuming a
power law with photon index $\Gamma$=1.7, 
absorbed by the Galactic column density toward the PKS~0312--77 field
(N$_H$=8$\times 10^{20}$ cm$^{-2}$, Dickey \&
Lockman 1990), and weighted by the effective exposure time of
the different EPIC cameras.
The uncertainty in the derived fluxes is $<15$\% for 
$\Delta \Gamma=\pm0.5$.
The 2--10 keV fluxes range from $\sim 1 \times 10^{-14}$ to
4 $\times 10^{-13}$ erg cm$^{-2}$ s$^{-1}$.\\
The same detection algorithm was also run in the 0.5--2 keV energy
range in order to characterize the
average spectral properties of the sources in our sample
using the hardness--ratio technique.

\par\noindent
The XMM--{\it Newton} observation of the PKS~0312--77 field
has been already analyzed by Lumb et al.~(2001); 
the same dataset is included in 
the First XMM--{\it Newton} Serendipitous Source Catalogue 
(2003, Version 1.0.1), recently released by the XMM-Newton Survey
Science Centre\footnote{http://xmmssc-www.star.le.ac.uk/} (SSC). 
A detailed comparison of the three 
samples is discussed in Appendix A.

\subsection{{\it Chandra}}
\par\noindent
The PKS~0312--77 field was also observed by {\it Chandra} during a
PV observation in 1999, September 8, for a total exposure time of
$\sim 12$ ks.
The analysis of the six objects selected in the 2--10 keV band
based on preliminary calibration data and detection techniques has
been already reported by Fiore et al. 2000 (hereinafter F00).
Thanks to the unprecedented {\it Chandra} positional accuracy
($\sim 1''$), it was possible to unambiguously identify
all of the optical counterparts of these hard X--ray sources.
\par\noindent
An additional, almost simultaneous {\it Chandra} observation of the
PKS~0312--77
field was retrieved from the archive, combined with the previous one, and
analyzed using version 2.2 of the CXC software.
The high--background intervals were filtered out leaving about 24.7 ksec
of useful data. The {\tt WAVDETECT} algorithm (Freeman et al. 2002)
was run on the cleaned full band (0.5--8 keV) image setting a
false--positive threshold of 10$^{-7}$, which led to highly reliable
detections, as shown in the HDF--N field (e.g., Brandt et al. 2001).\\
Twenty--five out of the 35 XMM--{\it Newton} sources are within the
{\it Chandra} ACIS--I field--of--view (FOV); only two sources detected
in the XMM--{\it Newton} observation
were not detected by {\it Chandra}: one object falls in a CCD
gap, while the other is just below the adopted probability threshold.\\
A detailed analysis of the {\it Chandra} X--ray spectral properties
and the comparison with the XMM--{\it Newton} results is
prevented  by the well known calibration problems related to the 
relatively high temperature of the ACIS--I detector during 
the PV phase.
We made use of the extremely good positional accuracy provided
by {\it Chandra} which allowed us to unambiguously estimate the X--ray
centroid position for the 23 common sources (see Sect. 3).

\subsection{Optical Imaging}
\par\noindent
The optical imaging ($R$-Bessel filter) of the PKS~0312--77 field was
carried out using EFOSC2 (Patat 1999) at the 3.6m ESO telescope in La~Silla
during three different observing runs (periods 66-68).
Exposure times were typically of 5-10~min, with a typical seeing of
about 1.5$''$. We acquired 16 $R$-band frames in order to cover all the 
X--ray sources; each single image, with a pixel size of 0.32 arcsec 
and a FOV of $\sim$5.3$\times$5.3 square arcmin, was cross-matched 
to the USNO catalog\footnote{http://archive.eso.org/skycat/servers/usnoa}
(Monet et~al. 1998) and astrometrically calibrated using the package GAIA
(version 2.3-1 driven P.W. Draper from the Skycat software developed
by ESO).  We obtained a good astrometric solution for each frame,
with r.m.s. in each coordinate of about 0.1-0.2$''$. 
The images were reduced using standard techniques including de--bias,
flat--fielding, and fringing correction (if needed).
The photometric calibration was performed for each night
using the zero-point derived from the measured
instrumental magnitudes of standard stars and assuming the average
extinction reported in the Observatory web 
page\footnote{http://www.eso.org/observing/support.html}.\\
The optical source catalogue was created using the {\tt SExtractor}
software (Bertin \& Arnouts 1996).
Since the images have been obtained under different seeing conditions,
%
%
the limiting magnitude has been conservatively estimated in each frame as
the 3$\sigma$ (sky value) over 2.5 times the seeing area. Using this
definition, the limiting magnitudes range between 24.0 and 25.2,
mainly depending on the image seeing. \\

\subsection{Optical Spectroscopy}
\par\noindent
The spectroscopic follow--up observations of the optically bright population 
(17 sources with R$<22$) have been performed with the ESO 3.6m
telescope equipped with EFOSC2 during four different 
observing runs (Jan 2000 -- Nov 2001) 
in the framework of the identification program of the {\tt HELLAS2XMM} survey.
We used the EFOSC2 grism \#13 with a 1.5~arcsec slit, which yields a
dispersion of about 2.8 \AA\ per pixel and provides a good spectral
coverage up to 9000 \AA. The exposure times vary between 600 and
2400~s, depending on the target magnitude. 
The spectroscopic follow--up of the 11 sources with R=$22-24$ 
has been performed with the ESO VLT/UT2 telescope equipped with FORS1
during period 69.  
The grism 150I with a 1.3~arcsec slit was used, providing a dispersion of
5.4 \AA\ per pixel and a wide spectral domain ($\approx 3500\div
10000$\AA).
The FORS1 exposure times range from 480 up to 6300~s;
dithering of the targets along the slits was applied for the faintest objects
in order to optimally remove the fringing at wavelengths longer than
7500 \AA.
\par\noindent
All the spectroscopic data have been reduced using standard
IRAF\footnote{IRAF is distributed by the
 National Optical Astronomy Observatories, which is operated by the Association
 of Universities for Research in Astronomy, Inc, under cooperative agreement
 with the National Science Foundation.} routines. 
Bias exposures taken on each night were stacked, checked for
consistency with the overscan regions of spectroscopic frames, and subtracted
out. The bias--subtracted frames were then flat--fielded in a standard manner
using internal lamp flats obtained during the same run. The sky background
was removed by fitting a third--order polynomial along the spatial direction
in source free regions. 
In all the observing runs the wavelength calibration was made using
arc lamps (He--Ar for the EFOSC2 data, He--Ar--Hg for the FORS1 data)
and different spectroscopic standard stars 
were used for the flux calibration.

\subsection{Radio}
\par\noindent
A deep radio observation of the PKS~0312--77 field at 5 GHz 
was performed with the Australian Telescope Compact Array (ATCA) in
the 6-km configuration (maximum baseline length), with a synthesized
beam size (HPBW) of $\sim 2''$.
The data were collected in a 12-hours run on 2000,  
September 27. In order to improve the sensitivity by a factor of
$\sqrt{2}$, we used both ATCA receivers at 5 GHz, centered at 
4800 and 5824 MHz, respectively. 
The field was observed in the mosaic mode by cycling through a grid 
of 5 pointings on the sky, in order to yield a uniform 
noise over the area covered by the {\it Chandra}
data.\\
The data were analyzed with the software package {\tt MIRIAD}. 
Since the ATCA correlator provides a bandwidth of 128 MHz subdivided into
32 frequency channels (of 4 MHz each), in the data reduction we
used the multi-frequency synthesis algorithms which give the opportunity
of producing images with improved ($u,v$) coverage by combining accurately
the visibility of individual channels.  The division of the wide
passband into subchannels reduces the effects of bandwidth smearing.
Each bandpass was calibrated and cleaned separately to
produce two individual images that were combined together into a single 
mosaic at the end of the reduction phase. Self--calibration was used 
to make additional correction to the antenna gains and to improve the 
image quality. The final map  has uniform noise of
50 $\mu$Jy (1$\sigma$) over an area with a semicircular shape (due to
the odd numbers of pointings) with a radius of about 10 arcmin, surrounded
by an area where the noise increases for increasing distance from the
center. The accuracy on the radio position is of the order of $\sim 2''$
for the faintest objects.

\subsection{Near--infrared}
\pn
 Deep K$_{\rm s}$ observations of a small subsample (ten objects) of the hard X--ray
 sources detected in the PKS~0312 field have been obtained with
 the Infrared Spectrometer And Array Camera (ISAAC, see Moorwood et al. 1999)
 mounted on the ESO VLT-UT1 telescope, as a part of complementary
 programs of the {\tt HELLAS2XMM} survey.
 The observations have been collected in service mode 
 during relatively good seeing conditions
 ($<0.8''$). 
We used the {\it ISAAC} SW Imaging Mode, which gives a pixel scale of
 0.1484 arcsec/pixel and a FOV of 2.5$\times$2.5 square arcmin. 
The net exposure time was 36 min for each field and, after
 running {\it SExtractor}, we estimated a 50\% completeness at
 $K_s\approx 21$ by comparing our data with deeper surveys 
(e.g., Saracco et al. 2001; Bershady et al. 1998).
The data reduction has been performed in two steps:
 individual raw frames have been first corrected for bias and dark current, and
 flat-fielded using standard IRAF routines. For the sky subtraction and
 image co-adding we then used DIMSUM \footnote{Deep Infrared Mosaicing
 Software, developed by P.~Eisenhardt, M.~Dickinson, A.~Stanford and J.Ward, 
 and available at {\tt ftp://iraf.noao.edu/contrib/dimsumV2}} a contributed
 package of IRAF. \\
 The K--band imaging data are used only for what concern the
 identification of the X--ray counterparts (see Sect. 3).
 We refer to Mignoli et al. (in preparation) for a full discussion on data
 reduction and analysis techniques.
\section{X--ray sources identification}
\par\noindent
\pn
At first, we have accounted for the astrometric calibration 
of the X--ray image by looking for average displacement 
of bright type 1 AGN already identified in F00. 
We found an average shift of $\sim$ 2$''$ 
($\Delta$(Ra)=1.67$''$; $\Delta$(dec)=$-1.13''$), in agreement with
the findings reported in Lumb et al. (2001), where the 
astrometric calibration was done with respect to the position of 
the bright central target.
 We have also verified the effects of possible scale and
rotational offsets in the matched astrometric solution, that 
turned out to be negligible. 
The uncertainties in the determination of the X--ray positions are 
mainly ascribed to the XMM--{\it Newton} PSF, in particular at faint
X--ray fluxes where the statistical error in RA and DEC determination 
are expected to be in the range 1 - 2 arcsec 
(see Sect. 6.3 in the First XMM-Newton Serendipitous Source Catalogue: 1XMM, User
Guide to the Catalogue).\\
\begin{figure}[!b]
\includegraphics[width=9cm, height=9cm]{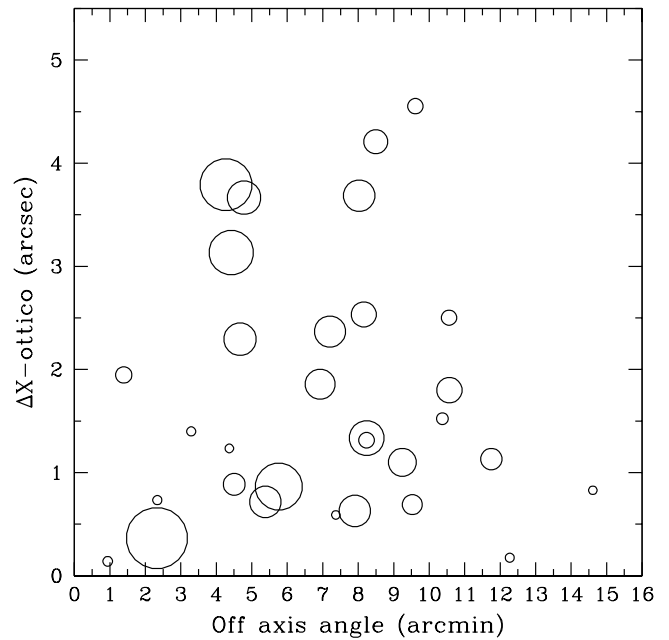}
\caption{Displacements between the 
optical and X--ray positions as a function of the off--axis angle. 
The size of the symbols increases as the 
X--ray flux decreases.}
\end{figure}                                                    
In order to accommodate any residual systematics in
the astrometric calibration of the EPIC images (see Barcons et
al. 2002) and to fully account for the PSF broadening in faint
sources (e.g., sources with $<$100 cts). 
we have searched for optical counterparts for all the X--ray sources 
within a conservative matching radius of 5 arcsec from the 
astrometrically corrected X--ray centroid.
Such a radius also represents the radius for which $\sim$ 95\%
of the XMM--{\it Newton} sources in the SSC catalogue are associated with 
USNO A.2 sources (see Fig. 7.5 in 
the First XMM-Newton Serendipitous Source Catalogue: 1XMM, User
Guide to the Catalogue).\\
Thirty--one X--ray sources have one or more optical counterparts
brighter than R$\sim$24.0 within the XMM--{\it Newton} error box, 
while for four objects there are no obvious counterparts down to 
the magnitude limits of the optical images (see Sect. 2.3).
The probability threshold adopted in the X--ray detection algorithm
corresponds to less than 1 spurious X--ray detection. 
We are then confident that also X--ray sources without optical counterparts
are real X--ray sources and in the following we address to these
objects as  blank fields.\\ 
In Fig. 1 we report the displacements between the 
optical and X--ray positions as a function of the off--axis angle. 
As shown by the size of the symbols (increasing as the 
X--ray flux decreases), a better X--ray--optical matching is achieved 
for the X--ray brightest sources, generally characterized by 
the sharpest PSF.
For these sources the displacements from the claimed optical
counterparts are $<2''$ even at large offaxis angles ($>12^{\prime}$). 
The average displacements ($\sim$~2 arcsec) of the X--ray faintest sources 
are consistent with those expected at these flux levels 
(see above). We finally note that the residual astrometric differences 
between the X--ray and optical positions do not show any clear trend
with the off--axis angle between 4 and 11 arcmin, where the bulk of the
sources are detected.\\ 
For each of the 29 XMM--{\it Newton} sources covered by radio observations
we also searched for radio sources within the X--ray error box (5$''$ radius). 
We found 5 X--ray/radio associations, while for the remaining 
24 X--ray sources we report the 3$\sigma$ upper limit (see Table
1).

\subsection{Confusion problems}
\pn 
\begin{figure}[!t]
\includegraphics[width=8.9cm,height=5cm]{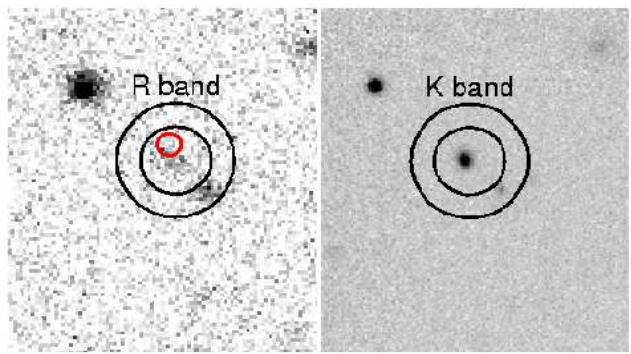}
\caption{({\it left panel}) R--band image of source 127. 
The small 1$''$ circle represents the position of the source 
detected by {\it Chandra};
the 3$''$ and 5$''$ large circles are centered on the XMM--{\it Newton} position;
({\it right panel}) K--band image for source 127. 
See Sect. 5.1 for details.}
\end{figure}
The additional {\it Chandra} data of the PKS~0312--77 field
allowed us to quantitatively investigate confusion problems
of X--ray sources, i.e. when the measured
X--ray emission is indeed originated from two or more X--ray sources 
at a distance comparable to the XMM--{\it Newton} PSF.
This could in principle be the case for four out of 31 sources, for which
two candidate optical counterparts fall within the XMM--{\it Newton} 
error circle (sources 20, 22, 127, and 18). 
Only in one case (source 18, also known as P3 [F00]) two objects are clearly resolved by {\it Chandra}
within the XMM--{\it Newton} detection (see Fig. 1 in Comastri et
al. 2002a); the X--ray flux of the faintest source is 
only about 10\% of the claimed counterpart, suggesting that most
of the XMM--{\it Newton} flux belongs to P3. 
The subarcsec positional accuracy of {\it Chandra} allows 
to unambiguously identify the correct optical counterparts of the 
remaining three sources and to exclude that two individual
X--ray sources contribute to the measured XMM--{\it Newton} flux
(see Fig. 2 for an example).\\
\noindent
\begin{figure}[!b]
\includegraphics[width=9cm,height=4.5cm]{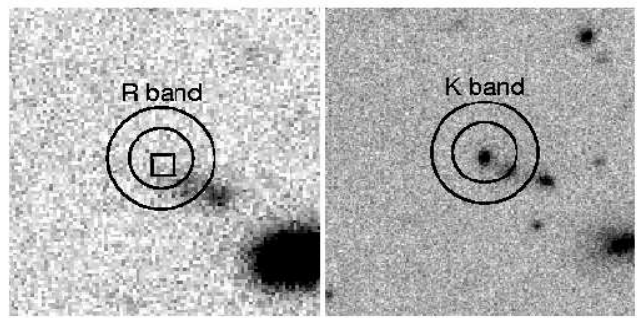}
\caption{({\it left panel}) R--band image of source 35. 
The small 1.5$''$ width box represents the radio detection position,
the 3$''$ and 5$''$ large circles are centered on the XMM--{\it
Newton} position; ({\it right panel}) K--band image for source 35.
See Sect. 3.1 for details.}  
\end{figure}
The multiwavelength coverage of our survey turned out to be extremely useful 
to investigate confusion problems also for the sources 
for which {\it Chandra} data are not available. 
The closest optical counterpart of source 35 (R=22), 
at a distance of $\sim 4''$ from the X--ray centroid,
has been identified with a Broad Line AGN at z=1.272; 
given the AGN surface density at these faint 
optical magnitudes ($\sim$100 deg$^{-2}$ at R=22, assuming a B-R color
of 0.6; Koo \& Kron 1988), the corresponding probability 
that the source lies entirely by chance in the XMM--{\it Newton} error box 
is 6.0$\times 10^{-4}$. 
However, a relatively bright ($S_{5 GHz}\sim 0.4$ mJy) radio source, 
associated with an optically blank field (R$>24.6$, see Fig. 3)
is almost coincident with the X--ray centroid.
A bright (K=18.5) source clearly emerges in the infrared band 
at the position of the radio source;   
the $R-K$ color $>5.1$ places this source among the 
Extremely Red Object (ERO; $R-K>5.0$) population.
Given the EROs surface density at K$<18.5$
($\sim$800 deg$^{-2}$, Daddi et al. 2000) and the fraction of
radio--emitters EROs at this level ($< 10$\%, see, e.g., Smail et
al. 2002; Roche et al.~2003) the corresponding probability 
that a radio emitting ERO lies entirely by chance in the XMM--{\it
Newton} error box is $< 5.0\times 10^{-4}$.
It is then likely that both the Broad Line AGN and the ERO 
contribute to the measured X--ray emission, although on the basis
of the chance coincidence argument we are not able to definitely 
disentangle the 
contribution of each source; therefore, we have associated half of the 
X--ray flux to the Broad Line AGN and half to the ERO.

We also note that the above described uncertainty in the 
source classification cannot be solved making use of the 
X--ray to optical flux ratio. Indeed even changing the X--ray flux by a factor
two the rest--frame $f_X/f_{opt}$ would be consistent with the 
typical values observed for broad line AGN and EROs respectively
(in the latter hypothesis the redshift has been estimated from the 
K--z relation of radiogalaxies).

\subsection{Sources at distance $>3''$}
\pn
Twenty--five out of 31 sources in the present sample
show optical counterparts within a 3$''$  radius from the XMM--{\it
Newton} centroid. 
The probability P of chance coincidence is in all cases, but one (see
Table 1) lower than 0.01 ($P=1-e^{-\pi r^{2}n(m)}$, where 
{\it n(m)} is derived from the number magnitude relation 
of field galaxies reported by Pozzetti \& Madau (2000)),
strongly supporting the reliability of our identifications.
\begin{figure}[!t]
\includegraphics[width=9cm,height=4.5cm]{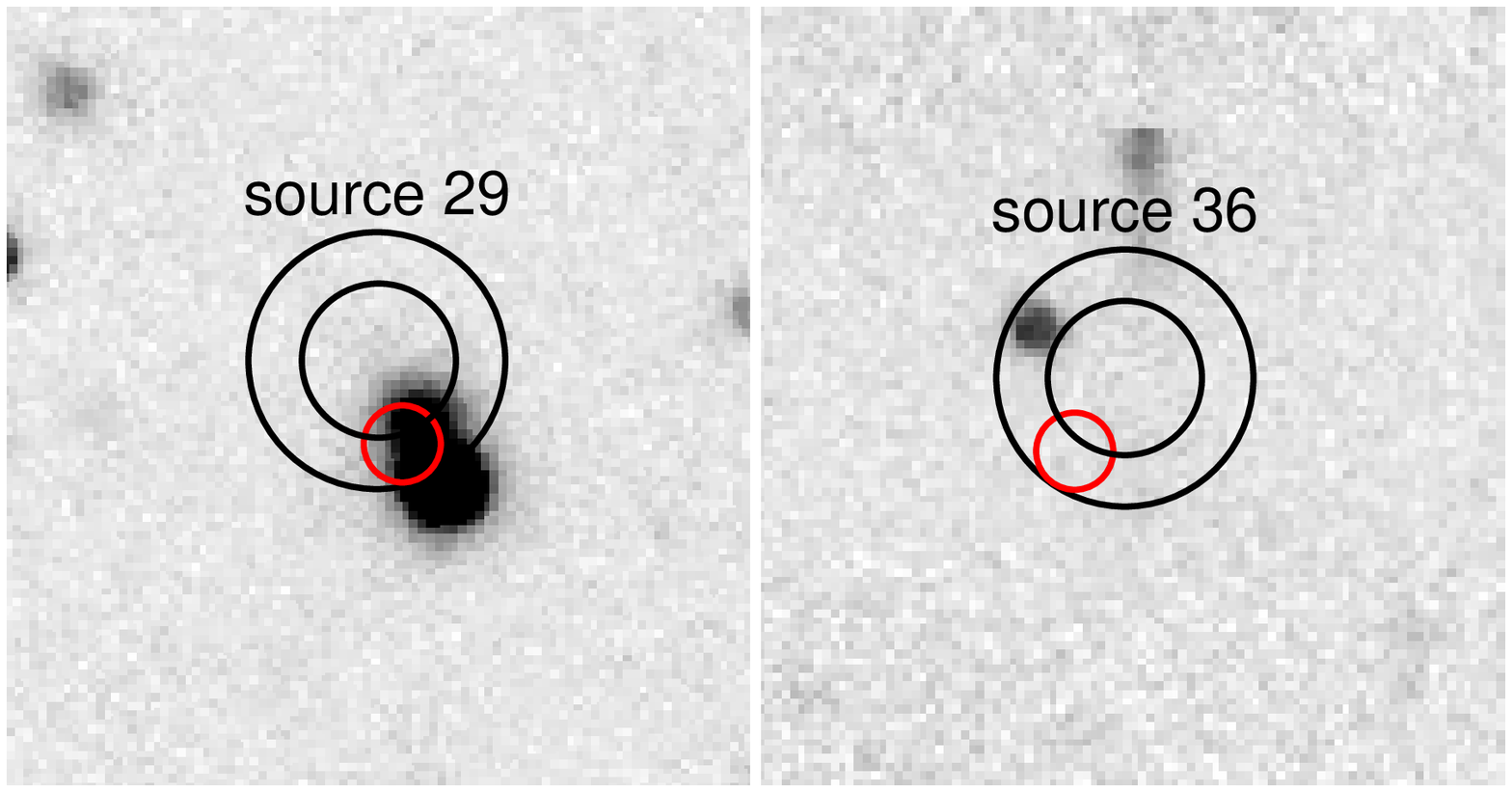}
\includegraphics[width=9cm,height=4.5cm]{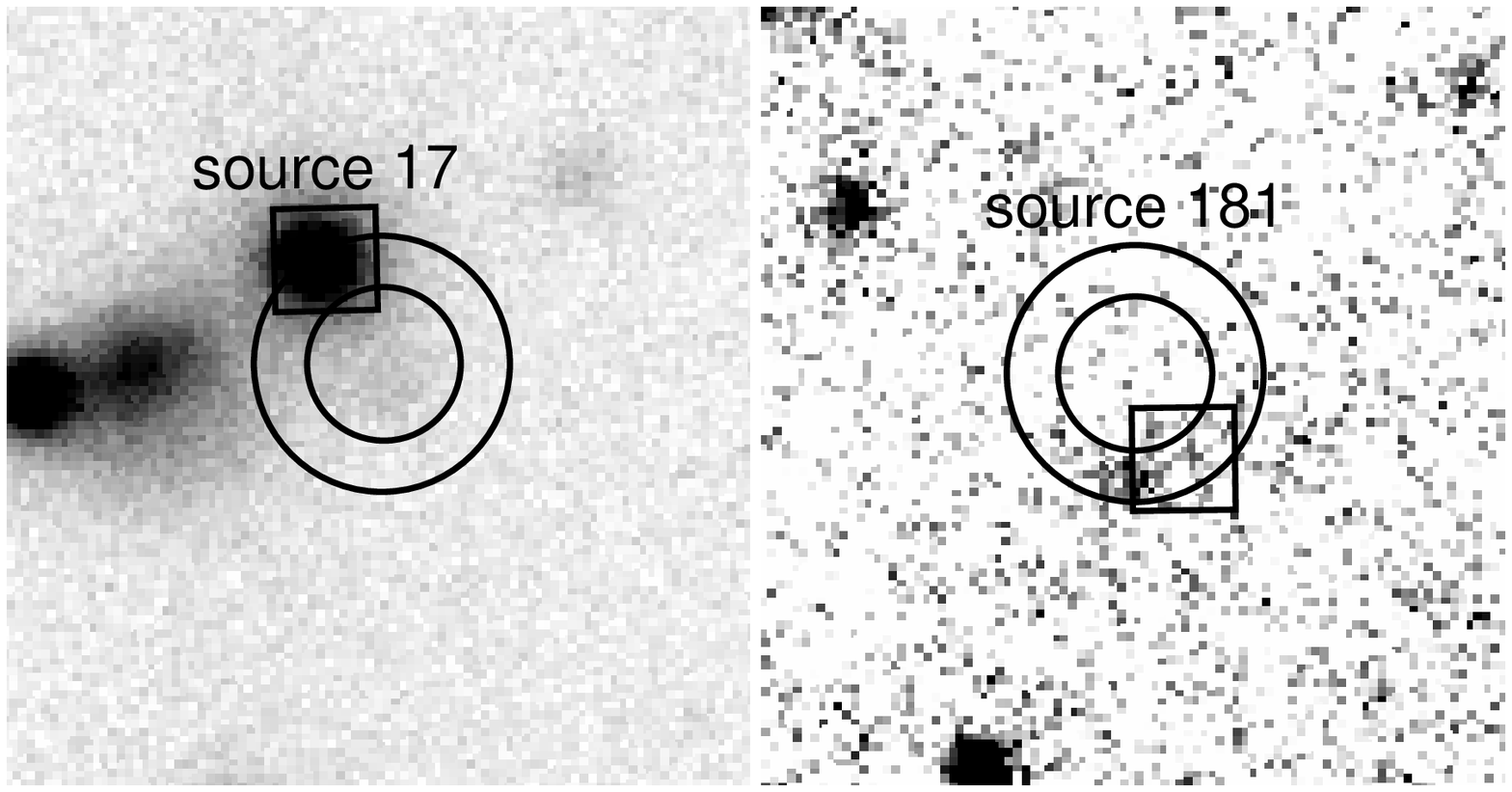}
\caption{EFOSC2 R--band images of 4 sources at a distance$>3''$ from the
XMM--{\it Newton} centroid. 
The 3$''$ and 5$''$ large circles are centered on the XMM--{\it
Newton} position. 
({\it top panels}) sources 29 and 36: the small circles refer to the {\it
Chandra} centroid and indicate a radius of 1.5$''$; 
({\it bottom left panel}) source 17 and 181: the squares refer to   
the radio centroids (4$''$ width). 
See Sect. 3.2 for details.}
\end{figure}
The {\it Chandra} and radio detections turned out to be extremely 
useful to identify five out of the six remaining sources for which
the closest optical counterpart lies between 3 and 5 arcsec from the
XMM--{\it Newton} centroid (Fig. 4): 
\begin{itemize}
\item[$\bullet$] the optical counterpart of source 29 
($r\simeq3''$) is a galaxy very close to a bright F/G star; it lies at
$\sim 1.5''$ from a {\it Chandra} source and it is most likely the
correct identification. 
\item[$\bullet$] a {\it Chandra} source within the XMM--{\it Newton}
error box of source 36 lies at $\sim 4.8''$ from the possible optical
counterpart which therefore cannot be the correct identification (see
Fig. 4). 
In the following we refer to this source as an optically blank
field.
\item[$\bullet$] 
the radio detection of the optical counterpart of source 17 ($r=4.5''$) 
strongly suggests that the identification is secure;
\item[$\bullet$] a faint radio source, detected at the 
3.5$\sigma$ level at $\sim 3''$ from the nearest optical 
counterpart of source 181 is likely to be the correct identification
given the  uncertainty on the radio position ($2-3''$)
\item[$\bullet$] source 35 has been already discussed
in the previous subsection. 
\end{itemize}

\noindent
The remaining object (source 66) is associated 
with a faint optical counterpart (R=23.1), classified as a Broad Line
AGN, at a distance of 3.7$''$. Given the AGN surface density at these faint 
magnitudes ($\sim$300 deg$^{-2}$, Mignoli et al.~2002)
the probability of chance coincidence is $1.8\times 10^{-3}$;
we are therefore confident that the identification is correct.\\ 
\par\noindent
Summarizing, we have securely identified 29 X--ray sources 
with a unique optical counterpart; 
in one case (source 35, see Sect. 3.1) we cannot provide an unique
identification.
Five X--ray sources turned out to be optically blank fields at the limit 
of our images. \\
The relative shifts between X--ray and optical centroids are plotted
in Fig. 5 ({\it left panel}) and also reported in Table 1. 
%
\begin{figure}[!t]
\includegraphics[width=9cm, height=9cm]{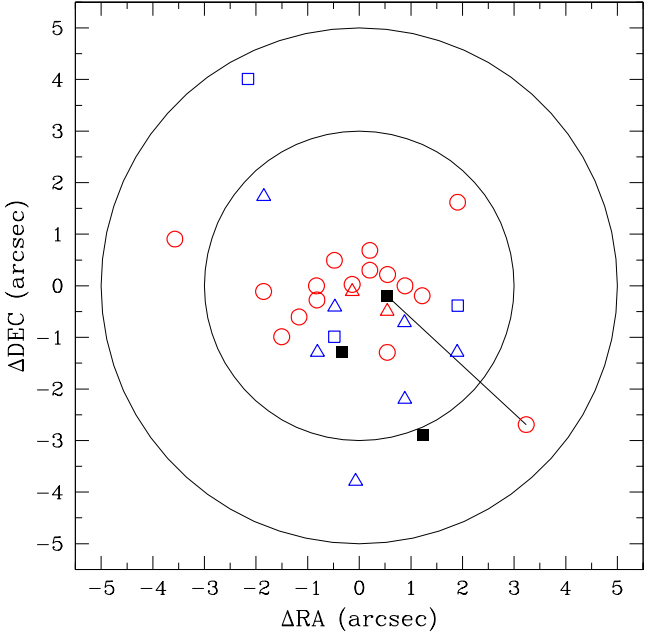}
\caption{Displacement between optical and X--ray positions of the 
proposed optical counterparts of the X--ray sources 
in the PKS~0312--77 field, 
after the astrometric calibration. Symbols are as 
follows: open circles are Broad Line AGNs, 
open triangles are Narrow Emission--line objects (Narrow Line AGNs
and ELGs), 
empty squares are galaxies, and filled squares are spectroscopically
unidentified sources.  
The solid line connects the two possible counterparts of source 35
(see Sect. 3.1).
The circles refer to the XMM--{\it Newton} error circles of 3'' and
5'', respectively.}   
\end{figure}

\subsection{Spectroscopic breakdown}
\pn
We have obtained good quality optical spectra for 28 out of 30
objects with R$<24$.
The optical counterparts have been classified into three broad categories:
\begin{itemize}                             
\item{\bf Broad Line AGN (BL AGN):} sixteen objects 
having broad emission lines (FWHM $\simgt 2000$ km s$^{-1}$);
\item{\bf Narrow Emission Line Galaxies (NELGs):} nine objects
including both narrow--line Type 2 AGN with FWHM $\simlt 2000$ km s$^{-1}$
or high-ionization state emission lines, and
extragalactic sources without obvious AGN features in their optical 
spectra but with the presence of at least one, strong emission feature
(Emission Line Galaxies, ELGs).
\item{\bf Normal galaxies} three objects showing a red continuum and
an absorption line optical spectrum, typical of an early--type
elliptical galaxy.
\end{itemize}

\section{Radio properties}

\begin{figure}[!t]
\includegraphics[width=8cm,height=8cm]{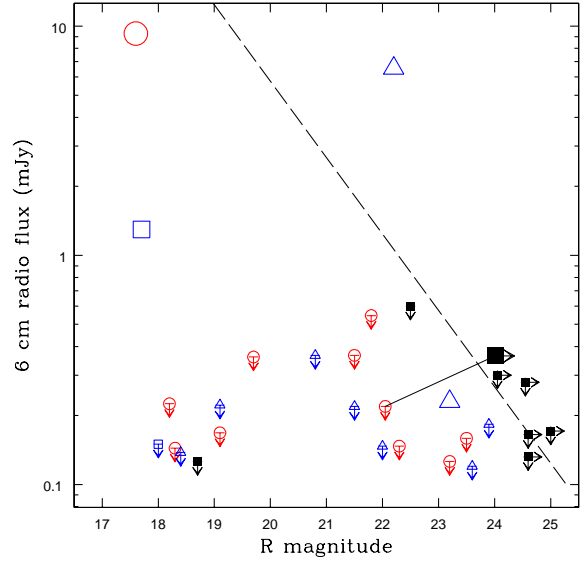}
\caption{The 5 GHz flux vs. R--band magnitude for the 29 sources 
with radio coverage. The dashed line represents the
locus of $\alpha_{ro}$=0.35 (see text for details). 
Symbols are as in Fig. 5. The solid line connects the two possible
counterparts of source 35 (see Sect. 3.1).
Enlarged symbols represent 
radio detections.}
\end{figure}
We have found five X--ray/radio associations out of 29 sources
with radio coverage. 
The 5 GHz flux versus the R band magnitude is reported in Fig. 6.
The dot--dashed line shows the value of $\alpha_{ro}$=0.35, often used to separate
Radio Loud (RL) and Radio Quiet (RQ) objects (see Ciliegi et al. 2003 for details).
Three out of the 5 X--ray/radio associations were clearly
detected at fluxes higher than 1 mJy. 
It is interesting to note that these radio--bright sources show
heterogeneous optical properties.
Only one object (source 16), spectroscopically identified with an ELG
at z=0.84, is classified as RL according to its $\alpha_{ro}$.  
The high--level of radio emission and its hard X--ray color (see Sect. 5.2) 
allow us to classify it as a Narrow Line
Radio Galaxy, the only flavor of obscured Type 2 AGN which has been extensively
studied at all redshifts and over a wide range of luminosities
(e.g. McCarthy 1993).   
The brightest radio source (S$_{5 GHz}=9.3$ mJy) is source 2, 
spectroscopically identified with a BL AGN (see also F00). 
Even if this object would not be classified as a RL source on the basis
of the observed $\alpha_{ro}$, the 5 GHz luminosity of $\sim 10^{26}$
Watt Hz$^{-1}$ is typical of RL quasars ($>10^{24}$ Watt Hz$^{-1}$).
Finally, a detailed discussion on the nature of the third bright object, 
an absorption line galaxy (source 17), is postponed to Sect. 5.3.\\
At fluxes below 1 mJy, we have detected radio emission in two 
sources (source 35 and source 181). 
\\
As discussed in Sect. 3.1, a radio source is almost coincident with
the XMM--{\it Newton} centroid of source \# 35.
On the basis of its $\alpha_{ro}$ value, this object 
would be classified as a RL object. 
If the K--z relation (Jarvis et al. 2001) of powerful radio galaxy
holds also at the mJy flux level, its bright K--band emission (K$\sim
18.5$, see Fig. 3) suggests that this object could be a high--redshift
(z=1.5-2) galaxy.\\
Finally, source 181 is classified as an ELG on the basis
of the optical spectrum.\\
Given the small area sampled and taking into account the radio coverage, 
the present X--ray sample is not suitable for a reliable estimate of the RL 
fraction among X--ray selected samples and a detailed analysis based on
a larger statistic is deferred to a forthcoming paper (Ciliegi et al.,
in preparation). 
\begin{figure*}[!ht]
\includegraphics[width=8cm,height=8cm]{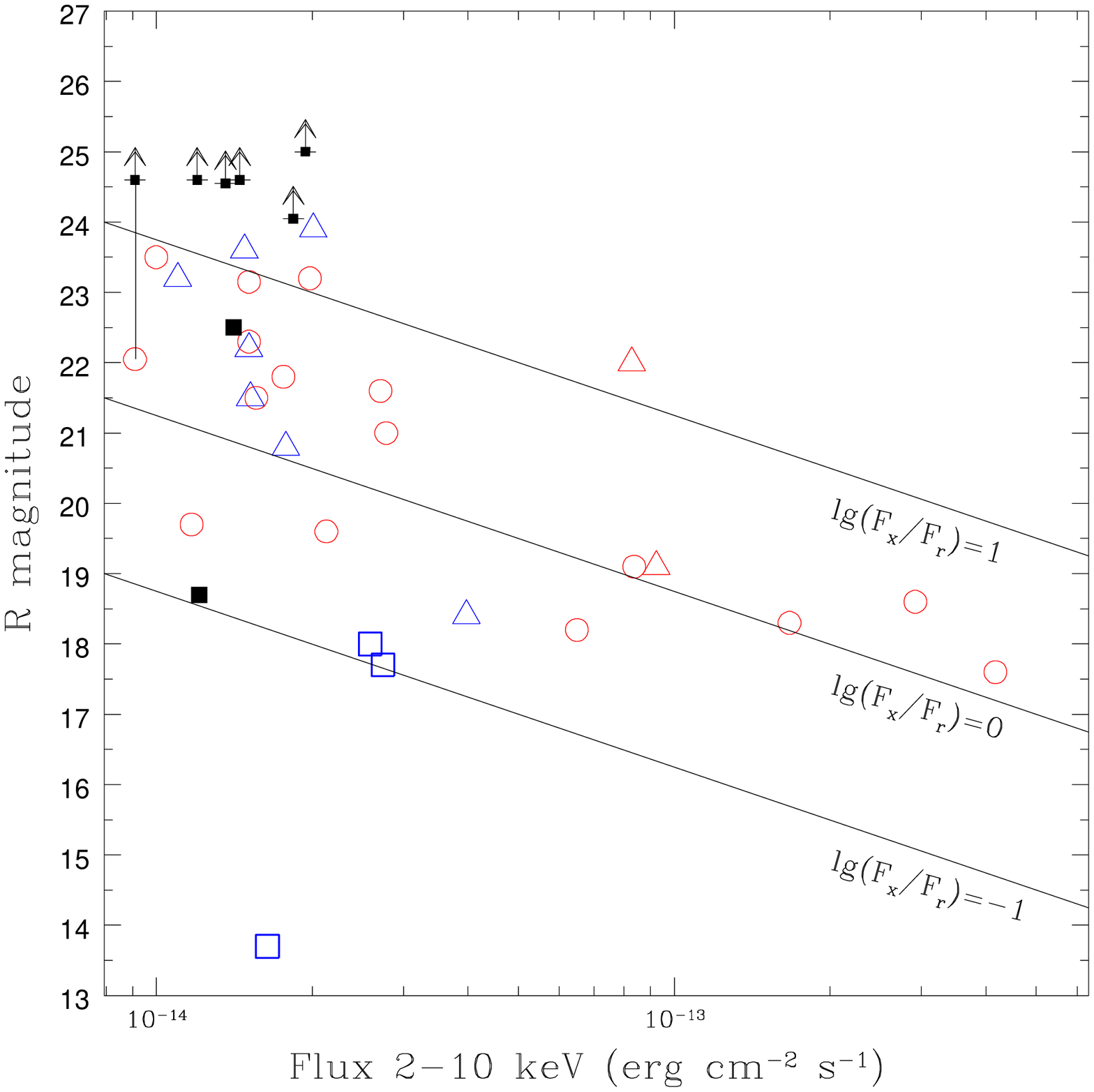}
\includegraphics[width=8cm,height=8cm]{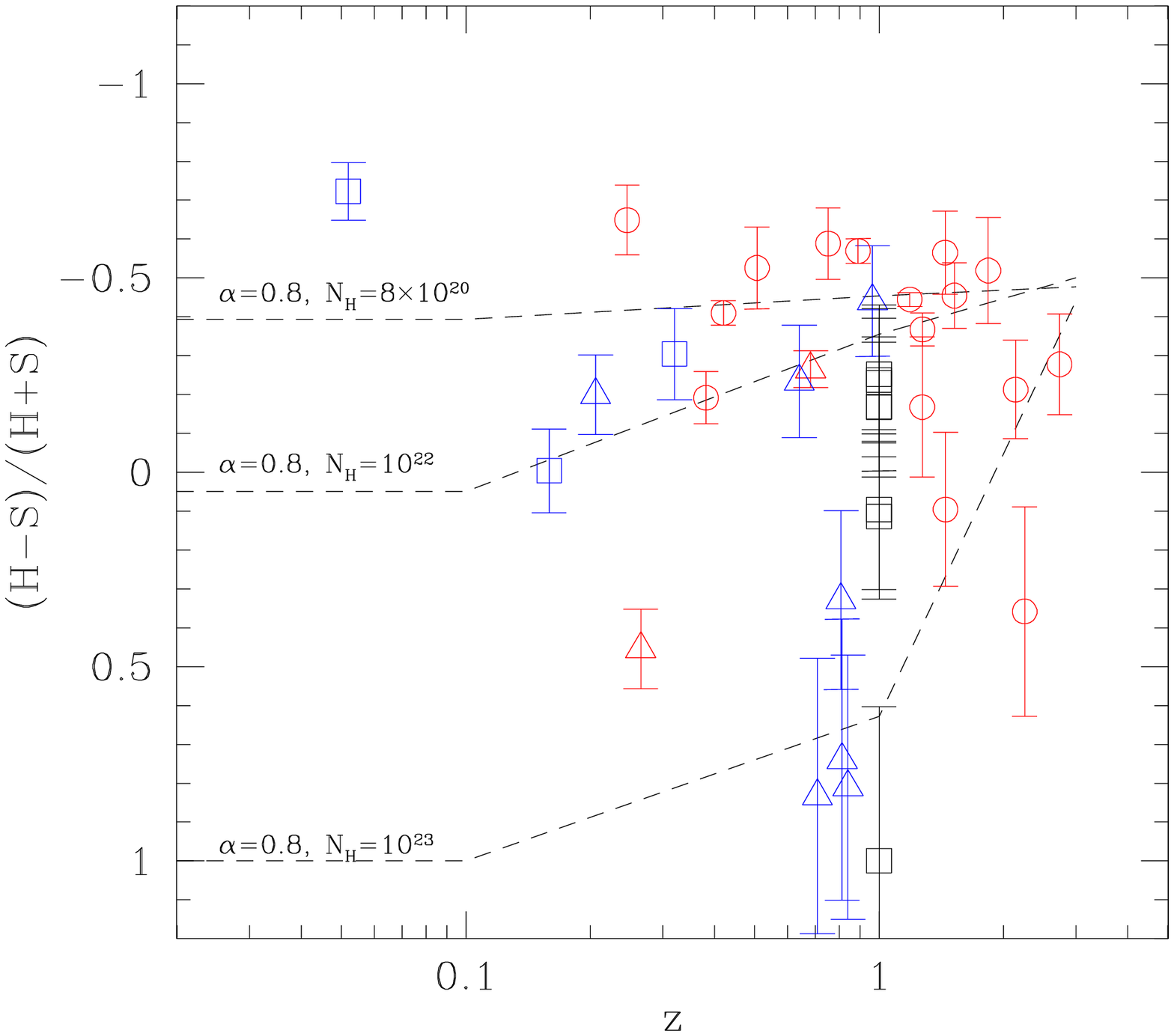}
\caption{({\it Left panel}) R--band magnitude vs. 2--10 keV X--ray
flux; the lines indicate the loci of  
constant X--ray--to--optical flux ratio (see Hornschemeier et al. 2000
for its definition). 
Symbols as in Fig. 5. 
The solid line connects the two possible counterparts of source 35
(see Sect. 3.1);
({\it Right panel}) Hardness ratio as a function of redshift.
Unidentified sources have been plotted at z=1.
For source 35, only the BL AGN counterpart has been plotted.
}
\end{figure*}

\section{Discussion}
\subsection{Broad--Line Objects}
\par\noindent
The 16 objects with broad optical emission lines have an average 
X--ray--to--optical flux ratio $<lg(F_{\rm x}/F_{\rm R})>$=0.32 with a
dispersion of 0.48 (see Fig. 7, {\it left panel}) which is typical of soft
X--ray selected Type 1 AGN (Lehmann et al. 2001). \\ 
Given the high spectroscopic completeness of the present
sample, it is possible to place some constraints on the {\it intrinsic} column
density (N$_{\rm H}$) directly from the HR vs. z diagram (Fig. 7, {\it right
panel}). 
While a population of BL AGNs with an {\it unobscured} X--ray
spectrum is present at all redshifts and spans a wide range in 
luminosities, several examples of {\it X--ray obscured}, BL AGNs 
seem to be present at high redshifts and X--ray luminosities. 
Taking the absorption column density obtained from the HR analysis 
at the face value, six out of the 13 BL AGNs with 
L$_{\rm X}>10^{44}$ erg s$^{-1}$ (Fig. 8) would have  
N$_{\rm H}> 10^{22}$ cm$^{-2}$, and two even larger than $10^{23}$ cm$^{-2}$ 
(Fig. 7, {\it right panel}).\\
Since the HR technique provides only a rough estimate of the
X--ray spectral properties, we have tried to constrain the 
intrinsic column density through proper spectral analysis.
A detailed description of the X--ray spectral fitting
results for all the 122 sources in the HELLAS2XMM sample is 
reported in Perola et al. (in preparation). 
Here we briefly discuss the results relevant for the purposes of
this paper, and in particular the 6 BL AGN
for which the HR analysis suggests the presence of significant
X--ray absorption. \\
The source spectrum and associate background and response files
for each of the 13 BL AGN (and for each XMM--{\it Newton} detector:
{\it pn} and {\it MOS}) 
have been extracted from the original event files 
using the standard procedures with version 5.4 of the SAS.
About half (six out of 13) of the sources are 
detected with a number of counts ($<$150) which does not allow
to adopt the standard $\chi^2$ minimization technique.
The use of the C--statistic, originally proposed  
by Cash (1979), is well suited  
to fit spectra with a few counts per bin in the limit of pure 
Poisson errors. 
In order to compare the results with those obtained with
the HR technique (Fig. 7, {\it left panel}), 
the power-law slope has been fixed to $\Gamma=1.8$.
The only parameters free to vary are the source normalization 
and the rest--frame intrinsic column density. 
There is no indication of intrinsic absorption for all of the 
seven sources which were considered as unobscured on the
basis of their HR. The same conclusion has been reached 
for two out of the six BL AGN classified 
as X--ray obscured from their HR (source \#35 and \#21),
though the absorption inferred from the HR is consistent
with the upper limit obtained from the spectral analysis.
For three of them (\#127, \#22 and \#66), 
the best-fit rest--frame N$_{\rm H}$ values are 17, 4.8, 4.7 $\times$ 
10$^{22}$ cm$^{-2}$, respectively, and larger than 
$\sim$ 10$^{22}$ cm$^{-2}$ at 90\% confidence level for all the 
sources.
These values are fully consistent with those obtained from the 
HR and provide further evidence toward the presence of 
substantial absorption in these objects.\\
The brightest X--ray obscured (N$_{\rm H}\sim$ 10$^{22}$ cm$^{-2}$ from the HR) 
BL AGN (source \# 7) 
has been detected with enough counts to constrain 
both the power-law slope and the absorption column density.
The best-fit parameters are fully consistent with the values
published by Piconcelli et al. (2002). The power law slope 
is rather flat ($\Gamma\simeq 1.4$) and the 90\% upper limit
on the intrinsic absorption is about 3.5 $\times$ 10$^{21}$ cm$^{-2}$.
If the power-law slope is fixed at $\Gamma$=1.8, the quality 
of the fit is slightly worse ($\Delta\chi^2$ = 3.1 for 1 d.o.f.) 
with a best-fit N$_{\rm H}$ =  3$\pm$2 $\times$ 10$^{21}$ cm$^{-2}$.
\pn
In order to assess the robustness of our procedure we have 
performed some tests.
First of all, we have changed the size of the local background 
regions to search for possible background fluctuations. 
Then we fit simultaneously the source plus background and 
background datasets linking together 
the model parameters for the background spectrum.
Finally for the ``brightest'' sources ($>$ 150 net counts) 
we have employed the standard $\chi^2$ fitting procedure.
In all the cases, the best fit parameters agree each other 
within the statistical errors.
\pn
It is concluded that the absorption column densities inferred 
from the observed HR are slightly overestimated. Nevertheless, 
there are evidences of substantial intrinsic obscuration
in three high-redshift sources spectroscopically classified as 
broad line AGN.
\begin{figure}[!h]
\includegraphics[width=10cm,height=9cm]{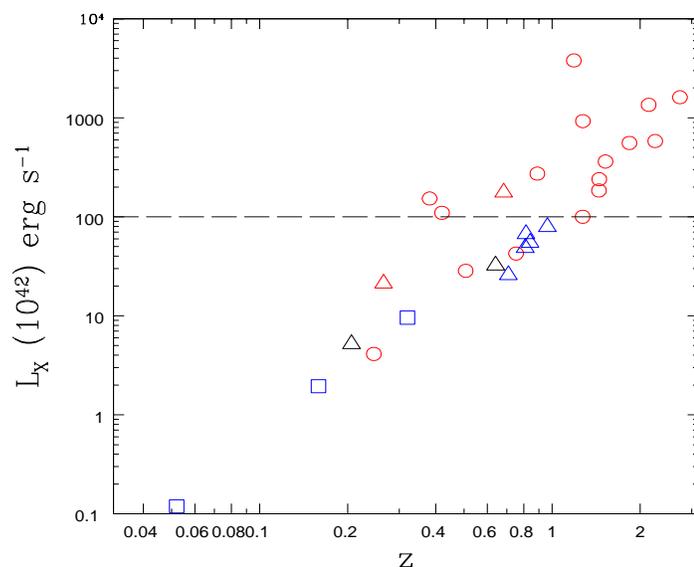}
\caption{The luminosity--redshift plane for all the identified sources in the
PKS~0312$-$77 field. Symbols are as in Fig. 5. 
X--ray luminosities are corrected for absorption. 
High(Low)--luminosity objects lie above(below) the dashed line at
L$_{\rm X}=10^{44}$ erg s$^{-1}$.} 
\end{figure}
The presence of broad optical emission lines and absorbed X--ray spectra
can be explained either by a dust--to--gas ratio significantly 
different from the Galactic one (Maiolino et al. 2001) or by  
geometrical effects such as a patchy X--ray absorbing medium 
on the same scale of the broad line region. 
\pn
The red optical--UV continua of some of the BL AGN in our sample suggest 
a dominant contribution of the host galaxy starlight
or an intrinsically absorbed continuum.
In this respect, the optical near--infrared properties of two objects 
in our sample are particularly interesting.\\
In source 7 (z=0.381), the presence of a broad H$\alpha$ line and the
relatively high X--ray luminosity ($\sim 2\times 10^{44}$ erg
s$^{-1}$), coupled with a red optical continuum dominated by the host galaxy 
starlight (as the presence of Ca H and K plus other absorption lines
clearly indicate, Fig. 9, {\it upper panel}), suggest to classify this
object as a {\it red quasar}. 
Other examples of low--redshift 
red quasars have been discovered both in X--ray
(Kim \& Elvis 1999; Vignali et al. 2000) and near--infrared 
surveys (Wilkes et al. 2002). 
It is puzzling that, although moderate  X--ray absorption 
(N$_{\rm H} \simeq$ 10$^{21-22}$ cm$^{-2}$) is often detected among 
red quasars, there is only marginal evidence of 
obscuration in source 7.  
The most convincing example of an optical spectrum which is dominated by 
the host galaxy starlight toward long wavelengths and by the 
active nucleus in the rest--frame UV has been 
recently reported by Page et al. (2003, see their Fig. 4).\\
Another very interesting object is the highly absorbed (N$_{\rm H}\sim10^{23}$
cm$^{-2}$) source 127 identified with a broad---line quasar at
z$=$2.251 on the basis of the MgII and CIII] lines (see Fig. 9, {\it
lower panel}). 
The other emission lines are narrow (FWHM$<$2000 km s$^{-1}$) and the
underlying continuum is  very red if compared to that of optically
selected BL AGNs (e.g., Brotherton et al. 2001). 
For this source additional K--band imaging has been obtained 
as part of a complementary program of the {\tt HELLAS2XMM} survey
(Mignoli et al., in preparation). 
The EFOSC2/R--band and VLT/K--band images are shown in Fig. 2.
The optical counterpart is associated with a bright near--infrared 
source (K= 18.4) having an optical to near--infrared 
color of R-K=5.1, considerably redder
than that of high--redshift quasars (R-K$\sim 2$), 
suggesting a dominant contribution from the host galaxy.
As far as the multiwavelength properties (continuum shape, 
luminosity, hard X--ray spectrum, and upper limit on the radio
emission) are concerned, 
this object is very similar to source N2\_25 in the ELAIS survey 
(Willott et al. 2003) which is indeed classified as a reddened
quasar at high redshift. \\
The examples discussed above highlight the need of a multiwavelength
coverage to properly classify and study the variety of continuum and
emission--line properties of the quasar population found in moderately
deep X--ray surveys. 
\begin{figure}
\includegraphics[width=8.5cm,height=6.5cm]{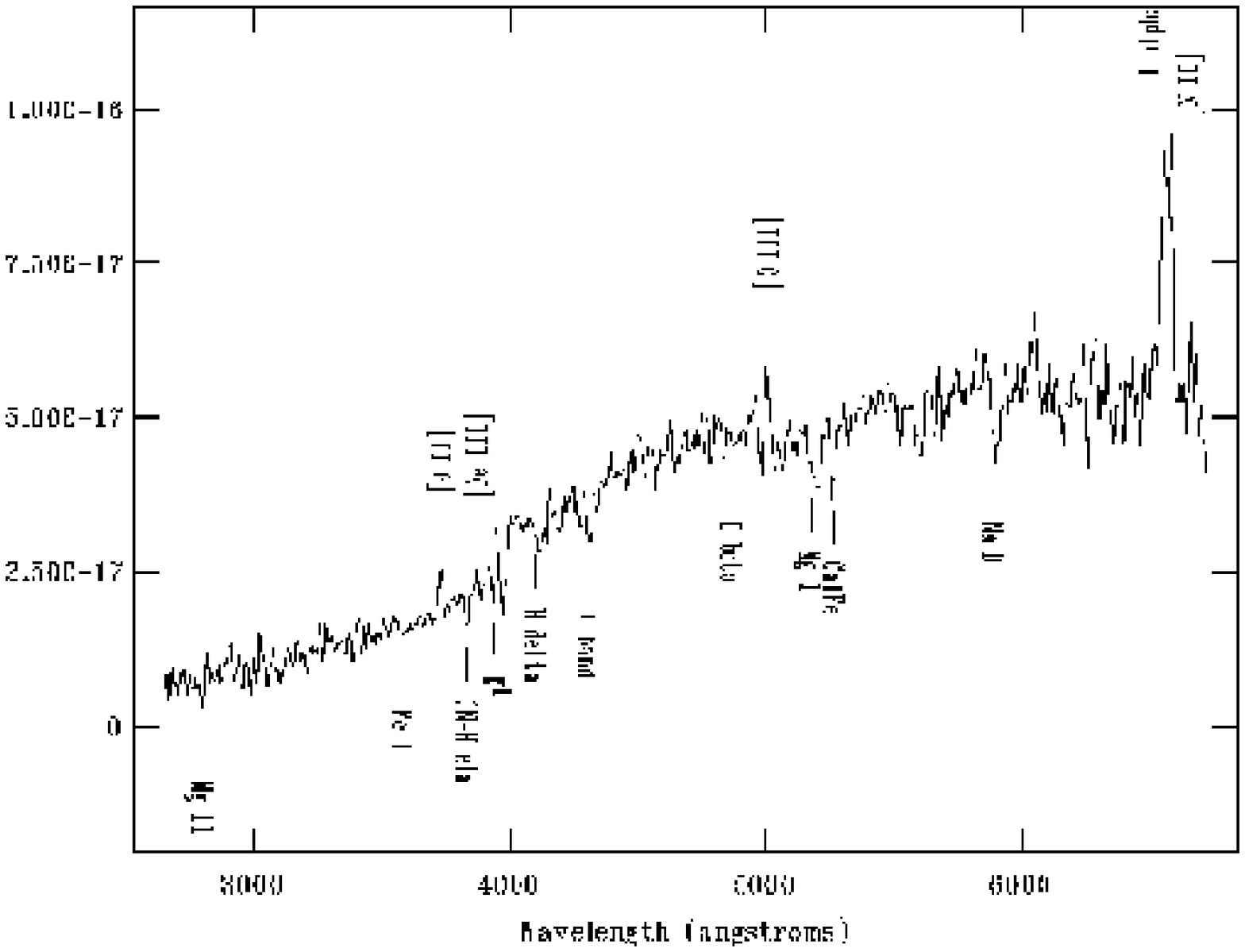}
\includegraphics[width=8.5cm,height=6.5cm]{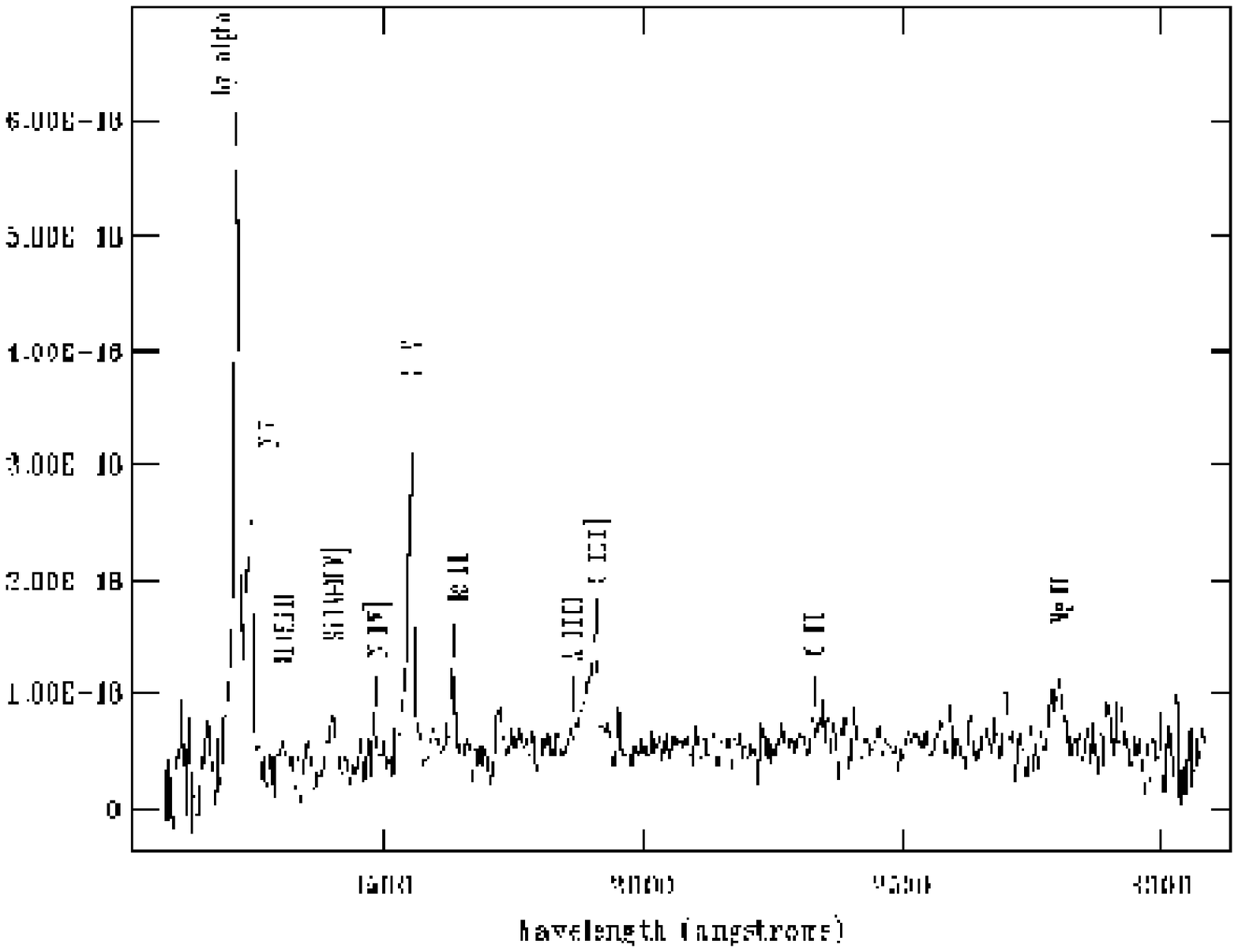}
\caption{The optical spectra of the {\it red quasar} at z=0.381
(source 7, upper panel) and the BL AGN/ERO at z=2.251 (source 127,
lower panel) discussed in Sect. 5.1} 
\end{figure}
\subsection{Narrow Emission Line Objects}
\par\noindent
Nine objects with narrow emission lines have been found.
On the basis of their optical lines ratios and intensities, 
two of them are securely classified as Type 2 AGN (sources 34 and 6) and one  
as a starburst galaxy (source 14), while for six objects 
(sources 20, 28, 89, 16, 181 and 116) we cannot provide a secure
classification (mostly because only one emission line is present in
the spectral range) and we refer to them as Emission--Line Galaxies
(ELGs).\\  
All of these sources lie at relatively low redshifts (z$<1$).
Except for source 14, spectroscopically identified with a starburst
galaxy, all the other objects in this class 
span a range of X--ray luminosities between 3$\times 10^{43}$ and
2$\times 10^{44}$ erg s$^{-1}$ (Fig. 8), typical of Seyfert 2 galaxies,
strongly suggesting the presence of an active nucleus even in the
objects where a more accurate optical classification is not possible.
Furthermore, in the context of AGN unified schemes, Type 2 AGN
are expected to be intrinsically absorbed sources.
Following the same procedure discussed in Sect. 5.1, we have estimated 
from the X--ray spectral analysis the absorbing column densities 
for these sources; both Type 2 AGN and four out of six objects among the ELG 
have N$_H$ values consistent with those obtained from the 
HR (larger than $10^{22}$ cm$^{-2}$ at the 90\% confidence level)
suggesting that the active nucleus is obscured both at X--ray and
optical frequencies.  
\subsection{Absorption Line Galaxies}
\par\noindent
One of the most surprising and unexpected findings of
deep and medium--deep {\it Chandra} and XMM--{\it Newton} surveys
is the discovery of X--ray bright sources (L$_{\rm X} \ga 10^{42}$ erg s$^{-1}$) 
in the nuclei of otherwise normal galaxies (F00; Mushotzky et al. 2000; 
Hornschemeier et al. 2000; Barger et al. 2001; Comastri et al. 2002b).\\ 
The three X--ray sources identified with normal absorption--line galaxies 
in the present sample show heterogeneous  properties. 
Two of them (source 18 and source 17) have X--ray--to--optical flux ratios 
($<lg(F_{\rm x}/F_{\rm R})>\sim -1$, see Fig. 7) which are marginally consistent 
with those of X--ray selected AGNs. 
The X--ray luminosity, almost two orders of magnitude higher 
than that expected on the basis of the L$_{\rm X}$--L$_{\rm B}$ correlation 
of early--type galaxies (Fabbiano et al.~1992), 
and the relatively hard X--ray colors strongly suggest that  
AGN activity is taking place in their nuclei.
The absorption--line optical spectrum 
of these X--ray Bright Optically Normal Galaxies (XBONGs)
have sometimes been explained if the nuclear light from a weak AGN 
is overshined by the stellar continuum of a relatively bright host 
galaxy (Moran et al. 2002; Severgnini et al. 2003);
However, a more complicate scenario for the two sources in our sample 
is suggested on the basis of multiwavelength data.  \\
A detailed study of 
source 18, which can be considered the prototype of
this class of objects, has been already reported in Comastri et
al. (2002a): the low level of radio emission and the broad--band
spectral energy distribution  favour the presence of an heavily
obscured, possibly Compton--thick (N$_{\rm H}>1.5\times 10^{24}$ cm$^{-2}$)
Seyfert--like nucleus. \\  
Alternatively, XBONGs could be the host galaxies of BL Lac objects;
if this were the case, deep radio observations would be the most useful
tool to test this possibility.
The relatively strong radio emission of source 17 (S$_{\rm 5
GHz}\sim 1.3$ mJy) and the relatively unobscured X--ray spectrum
(N$_{\rm H}\sim 10^{21}$ cm$^{-2}$) favours the BL Lac hypothesis. 
Furthermore, for source 17, the 5 GHz luminosity ($\sim 3\times 
10^{40}$ erg s$^{-1}$), which is considered a good indicator 
of the overall spectral energy distribution (see Fig. 7{\it (a)} in
Fossati et al. 1998), suggests that this object could be a rather
extreme example of a  high--energy peaked BL Lac (HBL; Ghisellini et
al. 1998; Costamante et al. 2001). This also explains the dominance of
the stellar continuum over the BL Lac non--thermal emission in the
optical band.
For comparison, a few examples of radio bright XBONGs have been
recently reported by Gunn et al. (2003) in the 13hr XMM--{\it Newton}
field. \\  
\par\noindent
The third object (source 8) is the softest source in the sample (HR=$-0.72$)
and is also characterized by the lowest X--ray--to--optical flux ratio
($<lg(F_{\rm x}/F_{\rm R})>\sim -2.5$). 
Deep R--band image revealed a complex morphology: 
the peak of the X--ray emission is almost coincident 
with the brightest nucleus of an interacting system at z$\sim$0.05 
containing at least three nuclei in a common envelope.
The total R--band magnitude of the system is R=13.7.
The optical spectra of the three nuclei are very similar,
showing an absorption--line spectrum
typical of early--type galaxies, without any emission line
except for a weak H$\alpha$ in the faintest one.
Both the optical and X--ray data favour thermal emission
from a small group of galaxies. The quality of the present XMM--{\it
Newton} observation, however, does not 
allow to distinguish between point--like and extended
X--ray emission. Unfortunately, the source lies at the edge of
the ACIS--I field in the {\it Chandra} observation, leaving this
issue unsettled.
Alternatively, if the X--ray emission were associated 
with a single nucleus, this object would be more similar to the 
previously discussed XBONGs.
\subsection{Unification schemes, absorbed X--ray sources, and the XRB}
\par\noindent
Recent multiwavelength programs of follow--up observations of
hard X--ray selected sources have already revealed a complex nature
for the hard X--ray source population (Barger et al. 2002, Giacconi et
al. 2002; Willott et al. 2003). They have also suggested the necessity
for substantial revision of the AGN unification models, which in their
simplest version (e.g., Antonucci 1993),  predict a one--to--one
relation between optical type 1 and X--ray unobscured sources,  and
between optical type 2 and X--ray obscured sources.
Despite the relatively low number of objects and the small area covered, 
the high spectroscopic completeness ($\sim 80\%$) and the
multiwavelength coverage of the PKS~0312--77 field allow to further
investigate this issue. 
As far as Seyfert luminosities (L$_{\rm X}\simlt10^{44}$ erg s$^{-1}$) are
concerned, our findings are in agreement with the prediction of
current unification models: the narrow--line AGNs are typically X--ray
obscured, while the BL AGNs are not.   \\
There seem to be, however, some hints for a departure from this 
simple scheme when quasar luminosities are considered; 
for example, the inferred column densities of three Type 1 QSOs are 
significantly higher than those estimated from the optical reddening
indicators assuming a Galactic extinction curve (see Sect. 5.1). \\
Other examples of  
X--ray absorbed, BL AGNs (both Seyfert and QSOs) have been discovered 
at relatively low redshifts among AGN selected from near--infrared surveys 
(2MASS AGN survey, Wilkes et al. 2002) and at brighter X--ray fluxes
in the ASCA and {\it Beppo}SAX surveys	(Akiyama et al. 2000; 
Fiore et al. 2001a,b; Comastri et al. 2001). 
Although it seems premature to claim that a population of 
X--ray obscured type 1 AGN has been discovered, it is interesting 
to note that, if a sizeable number of 
these objects will be found by more sensitive X--ray 
observations, they could have the same role
in contributing to the XRB of the so far elusive class of QSO2, for
which only a handful of objects are reported in literature (e.g.,
Stern et al. 2002; Norman et al. 2002; Mainieri et al. 2002).\\
For the five objects undetected in the optical images, no reliable
spectroscopic identification is possible even with 8m class
telescopes.  
These objects are characterized by an X--ray--to--optical (hereinafter X/O) flux ratio 
$> 10$, e.g. more than one order of magnitude larger than 
the value expected for optically selected AGNs (Lehmann et al. 2001).
Sources with such a high X/O ratio represent $\sim$ 25\% of the
present sample (9 objects, see Fig. 7) and their fraction seems to be
constant at lower fluxes in deep {\it Chandra}  and XMM--{\it Newton}
observations (see Fiore et al. 2003 for a detailed discussion). 
A large fraction of high X/O ratio sources is characterized by intrinsic 
column densities in excess of 10$^{22}$ cm$^{-2}$.
This is an indication that the majority of the high X/O are powered by an 
X--ray obscured AGN. 
As shown in Fiore et al 2003 (see their Sect. 5 for details),  
it has been proposed that a large fraction of these sources would lie
in the redshift range z=0.5$\div$2. 
If this were the case, at the fluxes of the present survey, 
high X/O sources would have an X--ray luminosity larger than 10$^{44}$
erg s$^{-1}$ and would contribute to increase the fraction of
high--luminosity, highly obscured sources closer to that predicted by
XRB models. \\ 
Moreover, preliminary results on VLT/ISAAC K--band imaging of
high X/O sources revealed that almost all of the optically blank
fields are associated with Extremely Red Objects (Mignoli et
al. 2003), further supporting the high--redshift nature for these
objects. 
Deep K--band spectroscopy would definitively test this hypothesis.
\section{Summary}
\par\noindent
The most important results obtained from the extensive 
multiwavelength coverage  of the {\tt HELLAS2XMM} field surrounding the 
radio--loud quasar PKS~0312--77 can be summarized as follows:

\begin{itemize}
\item[$\bullet$] We have detected 35 serendipitous hard X--ray sources
in a 30 Ks XMM--{\it Newton} observation in the field of the radio
loud--quasar PKS~0312--77. The X--ray sources span the flux range
between $1\div 40 \times 10^{-14}$ erg cm$^{-2}$ s$^{-1}$.

\item[$\bullet$] Thanks to the extremely good positional accuracy 
of {\it Chandra}, complemented by medium--deep radio and K--band observations, 
we have unambiguously identified the optical counterparts of 
$\sim$85\% of the hard X--ray selected sources. 
Taking into account the {\it Chandra} coverage of the
XMM--{\it Newton} field, coupled with radio and infrared data 
we have revealed confusion problems in 2 out of 35 sources
($\sim 6\%$) at an X--ray flux level of about 10$^{-14}$ \cgs.

\item[$\bullet$] At the relatively bright 2--10 keV fluxes sampled 
by our survey, the identified objects 
are characterized by a wide spread in their optical properties 
(both in the continuum shape and emission lines). 
The overall picture emerging from our study suggests 
that the optical appearance of hard X--ray selected AGN 
is different from what expected on the basis of the unified schemes, 
implying that classification schemes  
may not apply beyond the waveband in which they were made.

\item[$\bullet$] Optical spectra of X--ray absorbed
sources have revealed a few examples of high--redshift, high--luminosity 
objects optically classified as BL AGNs. 

\item[$\bullet$] The multiwavelength coverage of the three ``normal'',
X--ray bright galaxies 
made possible to further investigate the nature of this class of objects. 
Thanks to deep radio data, it was possible to tentatively identify a 
(H)BL Lac as the nuclear source of one object (source 17);  
the multiwavelength properties of source P3 have been already
extensively discussed (Comastri et al. 2002a) and favour a
Compton--thick scenario; the origin of the X--ray emission from the
third object is probably related to thermal radiation from an
interacting system.

\item[$\bullet$] Finally, about 25\%  of the objects in the
present sample have  X--ray--to--optical flux ratio greater than 10. 
Our results support the hypothesis that high X/O sources
are powered by obscured accretion at z$\sim$1. 
Deep near--infrared spectroscopy could provide 
a powerful tool to test this scenario.
\end{itemize}
\pn
The overall picture of the nature of hard X--ray selected 
sources has already approached a fairly complicate description;
it is clear that further and deeper observations over a broad 
range of frequencies would be helpful to better understand the content 
of hard X--ray sky.  

\begin{acknowledgements}
\par\noindent
We thank the entire {\it Chandra} team and in particular the CXC team
for the support received in the data analysis.
This paper used observations collected at the
Australian Telescope Compact Array (ATCA), which is founded by the
Commonwealth of Australia for operation as a National Facility by CSIRO.
The authors acknowledge partial support by ASI I/R/113/01 and I/R/073/01
contracts. 
MB and AC thank G. Zamorani for useful comments and discussion.
PS thanks the financial support from ASI I/R/037/01.
CV also thanks the NASA LTSA grant NAG5-8107 for financial support.
Last, but not least, comments by the referee greatly helped us to
improve the presentation of the results.
\end{acknowledgements}

\appendix
\section{Comparison with Lumb et al. 2001 and SSC sources}
 A list of serendipitous X--ray sources detected 
in the field of PKS~0312$-$77 was published by 
Lumb et al. (2001, hereinafter L01) and, more
recently, by the SSC consortium.
Here we try to investigate the differences in both the number of detected 
sources and their X-ray fluxes on the basis of the available information.\\
\begin{itemize}
\item {\sf L01:} 
Using a sliding--cell detection algorithm (the XMM--SAS task 
{\tt EBOXDETECT}) and a detection threshold of 5$\sigma$, 
52 sources were detected in the soft (0.5--2 keV) 
band; for 47 of these sources the 2--10 keV flux 
was also reported. \\
\item {\sf SSC:} 
A total of 142 sources were detected in at least one of the 
five different energy bands (0.2-0.5; 0.5-2; 2-4.5; 4.5-7.5; 7.5-12 keV) 
and/or in one of the three EPIC cameras, with a maximum likelihood parameter 
(ML, from the task {\tt EBOXDETECT}) larger than 8. 
\end{itemize}
\begin{figure*}[!t]
\includegraphics[width=8cm,height=8cm]{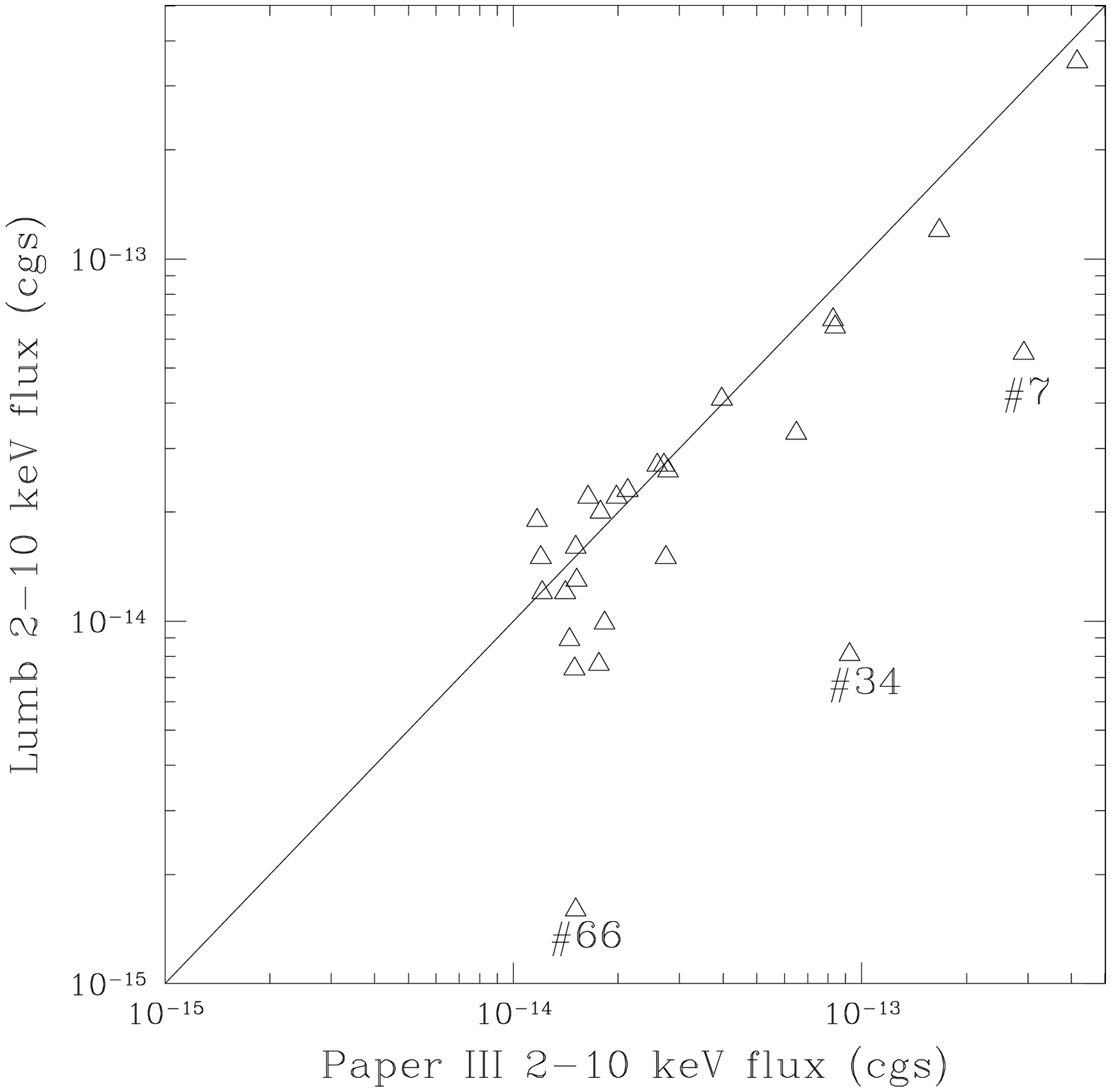}
\includegraphics[width=8cm,height=8cm]{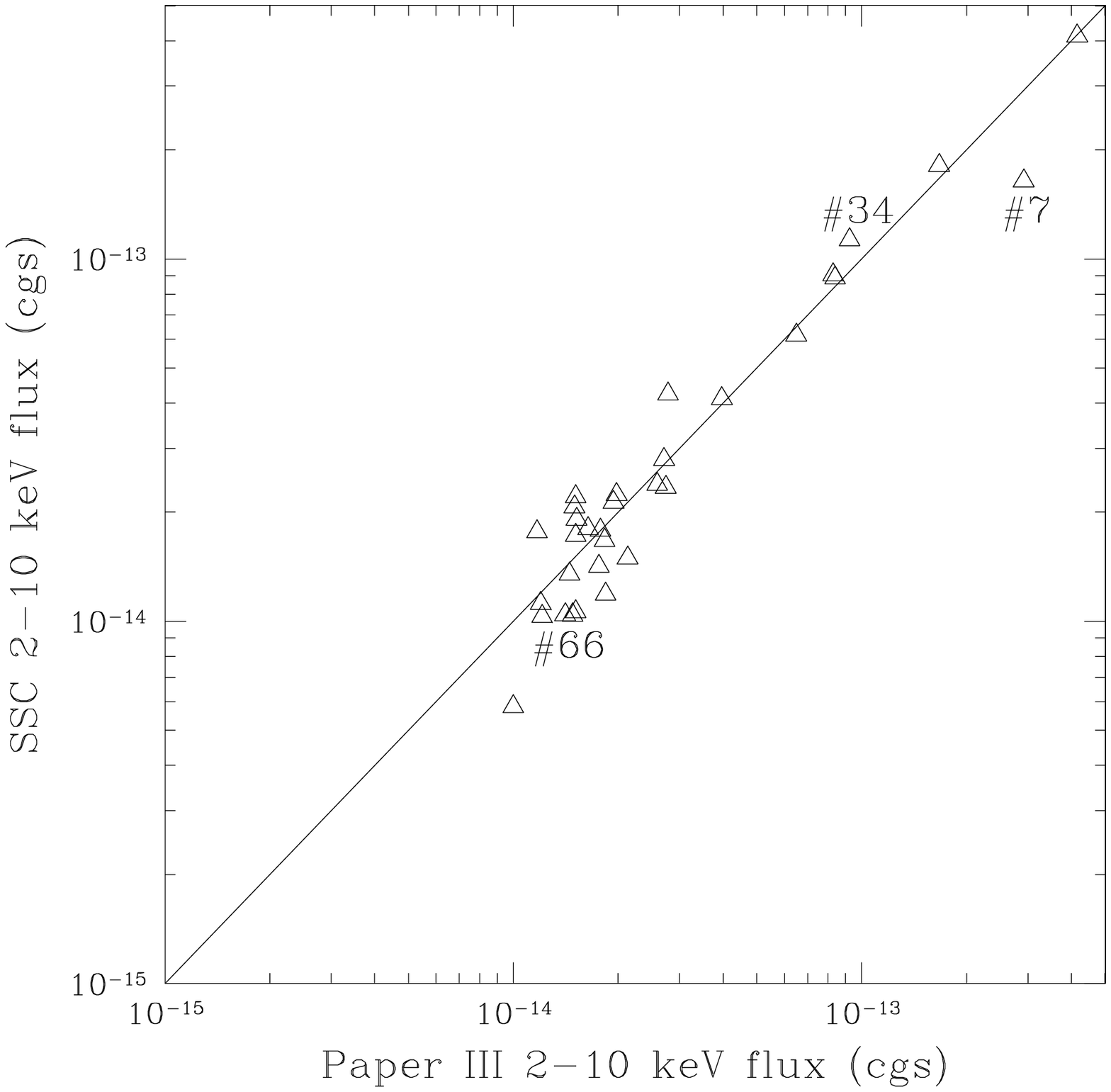}
\caption{({\it Left panel}) The comparison between the 2--10 keV
fluxes of our sample and those estimated by L01 for the 27 common
sources. The solid line represent the relation 1:1 between 
the two fluxes estimated. Three sources 
discussed in the Appendix are also shown in the plot.
{\it Right panel} As the {\it left panel}, but for the 32 common
sources in the SSC and our samples.  } 
\end{figure*}
%
The comparison between our sample, L01 sample and 
SSC sample is not straightforward due to the different detection 
algorithms, thresholds, band selection and, partly, to the different versions 
of the analysis software adopted in the data reduction. 
We also note that the L01 results have been obtained on the basis of 
preliminary calibrated data. \\
There is a significant difference in the total number of X-ray detected sources.
The origin of this discrepancy could be mainly ascribed 
to the hard X--ray selection of our sample. 
Twenty-seven of the 35 sources in our sample have been reported by L01; 
32 of the 35 sources have also been detected by the SSC 
when the detection in the 2-4.5 keV band (the energy range closest to the 2--10 keV 
band where the signal-to-noise ratio is maximized) is considered. 
We note that all of the sources in common between L01 and our sample are also 
included in the SSC sample. 
The five sources which belong to our sample and the SSC catalogue only 
(\# 127, 65, 16, 89, and 45 in Table~1) 
are detected with ML$>$12 in the SSC 2-4.5 keV band; therefore we 
consider these sources as highly reliable. 
The absence of these sources in L01 is likely due 
to their hard X--ray spectrum (all of them have HR$>$0; see Table~1). 
\\
Three sources are reported in our catalogue (\# 181, 501, and 116) but
not in the L01 and SSC ones; they are the hardest sources in our
sample, and appear robust X-ray detections by visual inspection.
The lack of these objects in the L01 and SSC sample is
most likely due to the hard X--ray selection.
\\
Four sources reported in L01 with fluxes brighter than our flux limit 
are not present in our sample, although they are included in the SSC
catalogue.  
Their non-detections in the SSC 2-4.5 keV band and their soft X-ray 
colors (quoted in Table~1 of L01) justify their absence in our 2-10
keV sample.  
\\
\\
We have also compared the flux measurements of individual objects  detected
above the limiting flux of our survey
($F_{2-10} \simeq$ 10$^{-14}$ erg cm$^{-2}$ s$^{-1}$)
which are in common in the three samples reported above.
Our investigation reveals that the L01 fluxes are typically lower than
ours by $\sim$~25--30\%  at fluxes $\simgt$~3 $\times$ 10$^{-14}$ erg
cm$^{-2}$ s$^{-1}$  (see Fig. A.1, {\it left panel}). 
Since the fluxes in the two samples have been computed assuming the 
same spectral model ($\Gamma=1.7$ and Galactic absorption), 
the differences are likely to be due to a different correction adopted 
for the encircled energy fraction. This is particularly pronounced for the 
bright sources detected at large off--axis angles (see, e.g., source \# 7). 
Two sources in the L01 sample deviate by more than one order of
magnitude from our flux measurements (\# 34 and \# 66); 
we think that this is likely due to typos in Table~1 of L01. 
\\
Conversely, there is a good agreement between our fluxes 
and the SSC ones (see Fig. A.1, {\it right panel}) for the 32 common sources. 
This result gives further support to the reliability of our flux estimates. 
The SSC 2-10 keV fluxes have been computed using the summed count rates in the 
2-4.5 and 4.5-7.5 keV bands extrapolated to the 2-10 keV band assuming 
the same spectral model adopted in this paper.

\input{h4366tab.tex}

\end{document}

%% file: h4366tab.tex
\begin{sidewaystable*}
\footnotesize 
\begin{tabular}{rrrcccrrrcll} 
\multicolumn{12}{l}{Multiwavelength properties 
of the X--ray sources detected in the XMM--{\it Newton}
observations}\\
\hline
ID    &  \multicolumn{2}{c}{XMM position} & 
$\Delta_{XMM-opt}$ & $\Delta_{Chandra-opt}$ &
{\it P} & S$_{5 Ghz}$   & R & 
F$_{2-10}$ & HR   & z & Class\\  
 & RA   & DEC    & [$''$] & [$''$] &   & [mJy]   &   &[10$^{-14}$ cgs]    
 & & \\  
\hline
  12 & 03 15 29.36 & $-$76 53 41.61 &  2.50  & \ldots$^{a}$  & $<0.01$ 
     & \ldots &   21.0 & 2.78$\pm  0.74$ & $-0.52\pm0.10$ & 0.507  & AGN 1 \\  
  20 & 03 14 16.76 & $-$76 55 59.51 &  2.37  & 1.42 &$<0.01$ 
     & $<0.219$&   21.5 & 1.52$\pm  0.47$ & $-0.44\pm0.14$ & 0.964 & ELG \\
  10 & 03 14 16.36 & $-$76 45 36.51 &  0.69  &	2.60$^{b}$	& $<0.01$
     & \ldots &   19.6 & 2.13$\pm  0.61$ & $-0.65\pm0.09$ &	0.247 & AGN 1 \\
  22 & 03 13 48.96 & $-$76 45 59.41 &  1.31  &	0.31	& $<0.01$ 
     & \ldots &   21.6 & 2.71$\pm  0.58$ & $-0.21\pm0.13$ & 2.140 & AGN 1 \\
  36 & 03 13 43.55 & $-$76 54 26.26 &  3.47  &	4.77	& 0.013	
     & $<0.165$&  $>$24.6 & 1.45$\pm  0.47$ & $-0.23\pm0.19$ & \ldots  & \ldots \\
  13 & 03 13 34.26 & $-$76 48 29.71 &  0.86  &	0.66	&$<0.01$ 
     & $<0.360$ &   19.7 & 1.17$\pm  0.38$ & $-0.56\pm0.11$ & 	1.446 & AGN 1\\
   3 & 03 13 14.66 & $-$76 55 55.81 &  1.23  &	1.04	& $<0.01$ 
     & $<0.150$&   18.3 & 16.72$\pm  1.03$ & $-0.41\pm0.03$ &	0.420  & AGN 1$^{\dag}$\\
   5 & 03 13 11.96 & $-$76 54 29.71 &  1.40  &     0.64	& $<0.01$ 
     & $<0.168$&   19.1 & 8.37$\pm  0.68$ & $-0.37\pm0.04$ &	1.274 & AGN 1$^{\dag}$\\
 127 & 03 12 57.96 & $-$76 51 20.31 &  0.36  &	0.55   & $<0.01$ 
     & $<0.159$&   23.5 & 1.00$\pm  0.29$ &  $0.36\pm0.27$ & 2.251  & AGN 1 \\
  65 & 03 12 52.16 & $-$77 00 59.51 &\ldots $^{*}$&\ldots $^{*}$ & \ldots 
     & $<0.300$&  $>$24.0 & 1.84$\pm  0.51$ & $0.11\pm0.21$ & \ldots  & \ldots \\
   6 & 03 12 53.96 & $-$76 54 14.51 &  0.73  &	0.70	& $<0.01$ 
     & $<0.150$&   22.0 & 8.28$\pm  0.66$ & $-0.26\pm0.05$ &	0.683  & AGN 2$^{\dag}$ \\
  18 & 03 12 39.26 & $-$76 51 32.61 &  1.94  &	1.29	& $<0.01$ 
     & $<0.150$&   18.0 & 2.59$\pm  0.39$ & $0.00\pm0.11$ & 0.159 & GAL$^{\dag}$\\
   8 & 03 12 31.16 & $-$76 43 24.01 &  1.10  &	0.62	& $<0.01$ 
     & \ldots &   13.7$^{e}$ & 1.64$\pm  0.52$ &  $-0.72\pm0.07$ & 	0.0517  & GAL/  \\
     &             &                &  2.90  &  1.49     &         
     &         &        &                 &  &   0.0537  & group?\\     
     &             &                &  2.16  &  2.15     &         
     &         &        &                 &  &  0.0517  & \\ 	
   4 & 03 12 09.16  & $-$76 52 13.01 &  0.14  &	0.60	& $<0.01$ 
     & $<0.225$&   18.2 & 6.49$\pm  0.56$ & $-0.57\pm0.03$ &	0.890 & AGN 1$^{\dag}$\\
  16 & 03 12 00.36 & $-$77 00 25.61 &  0.63  & 1.33	& $<0.01$ 
     & 6.545$\pm0.500$&   22.2 & 1.51$\pm  0.46$ & $0.81\pm0.34$ &  0.841 & ELG \\
  89 & 03 11 44.96 & $-$76 56 45.01 &  2.29  &	1.13	& 0.054 
     & $<0.150$&   23.6 & 1.48$\pm  0.37$ & $0.33\pm0.23$ & 0.809  & ELG \\
 181 & 03 11 35.96 & $-$76 55 55.81 & 3.79 & \ldots$^{c}$& 0.104 
     & 0.230$\pm0.050$& 23.2 &1.10$\pm0.32$& $0.83\pm0.35$ &  0.709  & ELG \\
  35 & 03 11 31.76 & $-$77 00 36.31 &  4.21  &\ldots $^{a}$&0.048 
     & $<0.360$ &  22.0   & 1.83$\pm0.53$& $-0.17\pm0.18$ & 1.272 & AGN 1\\
     &         &               & 0.60$^{h}$ &\ldots $^{a}$& \ldots
     & 0.364$\pm0.073$ & $>24.6$ & & &\ldots  &\ldots\\ 
\hline
\end{tabular} 
\end{sidewaystable*}
\newpage
\begin{sidewaystable*}
\footnotesize 
\begin{tabular}{rrrcccrrrcll} 
\multicolumn{12}{l}{Multiwavelength properties 
of the X--ray sources detected in the XMM--{\it Newton}
observations (continued)}\\
\hline
ID    &  \multicolumn{2}{c}{XMM position} & 
$\Delta_{XMM-opt}$ & $\Delta_{Chandra-opt}$ &
{\it P} & S$_{5 Ghz}$   & R & 
F$_{2-10}$ & HR   & z & Class\\  
 & RA   & DEC    & [$''$] & [$''$] &   & [mJy]   &   &[10$^{-14}$ cgs]    
 & & \\  
\hline
  66 & 03 11 28.16 & $-$76 45 16.31 &  3.68  & \ldots $^{d}$& 0.09 
     & \ldots &    23.1   & 1.51$\pm  0.46$ & $0.09\pm0.20$ & 1.449 & AGN 1 \\
  17 & 03 11 24.76 & $-$77 01 39.01 &  4.55  &\ldots  $^{a}$&$<0.01$
     & 1.298$\pm0.088$&   17.7 & 2.74$\pm  0.60$ & $-0.30\pm0.13$ & 	0.320 &  BL Lac \\
  31 & 03 11 13.86 & $-$76 53 59.11 &\ldots $^{*}$&\ldots $^{*}$&\ldots
     & $<0.150$&   $>24.6$ & 1.20$\pm  0.33$ & $-0.17\pm0.16$ & \ldots &\ldots \\
  29 & 03 11 13.36 & $-$76 54 31.11 &  3.13  &	1.57	& $<0.01$ 
     & $<0.150$&   18.7 & 1.21$\pm  0.33$ & $-0.24\pm0.16$ & \ldots & \ldots   \\
  11 & 03 11 12.76 & $-$76 47 01.91 &  1.86  &	1.50	& $<0.01$ 
     & $<0.366$&   21.5 & 1.56$\pm  0.49$ & $-0.59\pm0.09$ & 0.753      & AGN 1 \\
   9 & 03 11 05.56 & $-$76 51 58.01 &  0.89  &	0.13	& $<0.01$ 
     & $<0.150$&   23.2  & 1.98$\pm  0.38$ & $-0.45\pm0.08$ & 1.522   & AGN 1 \\
   7 & 03 10 49.96 & $-$76 39 04.01 &  0.83  &\ldots $^{a}$ & $<0.01$ 
     & \ldots &   18.6 & 29.2$\pm  3.10$ & $-0.19\pm0.07$ &	0.381 & AGN 1 \\
  21 & 03 10 49.76 & $-$76 53 16.71 &  0.72  &	0.24    & $<0.01$ 
     & $<0.150$&   22.3 & 1.51$\pm  0.36$ &  $-0.28\pm0.13$ &  2.736 & AGN 1 \\
  28 & 03 10 37.36 & $-$76 47 12.71 &  2.53  &	1.09	& $<0.01$ 
     & $<0.366$&   20.8 & 1.78$\pm  0.46$ & $-0.23\pm0.14$ &	0.641 &  ELG \\
  45 & 03 10 18.96 & $-$76 59 57.91 &\ldots $^{*}$&\ldots$^{a}$&\ldots
     & $<0.171$&   $>25.0$ & 1.94$\pm  0.57$ & $0.09\pm0.20$ & \ldots  &\ldots \\
   2 & 03 10 15.76 & $-$76 51 33.21 &  0.59  & 0.64	&$<0.01$ 
     & 9.284$\pm0.073$&   17.6 & 41.6$\pm  1.60$ & $-0.44\pm0.02$ & 1.187 & AGN 1$^{\dag}$ \\
 124 & 03 10 01.60 & $-$76 51 06.71  &  1.33  &  1.88    &$<0.01$ 
     & $<0.300$&   22.5 & 1.41$\pm  0.45$ & $-0.25\pm0.18$ &        & \ldots \\
 501 & 03 09 52.16 & $-$76 49 27.41 & \ldots $^{*}$ &	\ldots  $^{a}$&\ldots
     & $<0.279$&   $>$24.5    & 1.36$\pm  0.45$ & $1\pm0.40$ & \ldots & \ldots$^{f}$ \\ 
 14  & 03 09 51.16 & $-$76 58 24.71 &  1.52  &	\ldots $^{a}$ &$<0.01$ 
     & $<0.150$&   18.4 & 3.97$\pm  0.70 $ & $-0.20\pm0.10$ & 0.206 & Starburst \\
 24  & 03 09 31.66 & $-$76 48 45.01 &  1.80  &	\ldots $^{a}$ & $<0.01$ 
     & $<0.546$&   21.8 & 1.76$\pm  0.54$ & $-0.52\pm0.13$ & 1.838 & AGN 1 \\
 116 & 03 09 18.46 & $-$76 57 59.31 &  1.13  &	\ldots $^{a}$ & 0.017 
     & $<0.183$&   23.9 & 2.01$\pm  0.67$ & $0.74\pm0.36$ & 0.814$^{g}$ & ELG   \\
  34 & 03 09 12.06 & $-$76 58 25.91 &  0.17  &	\ldots $^{a}$ & $<0.01$ 
     & $<0.222$&   19.1 & 9.24$\pm  1.13$ & $0.45\pm0.10$ & 0.265 & AGN 2 \\
\hline
\end{tabular} 
\caption{$^{*}$ blank field;
$^{a}$ out of {\it Chandra} ACIS--I field; $^{b}$ at the edge of ACIS
field; $^{c}$ CCD gap; $^{d}$ not detected. 
$^{e}$: the R--band magnitude refer to the three nuclei altogether
(see \S 5.3);
$^{f}$: we note that a R=20.1 counterpart lies at 5.9$''$
from the XMM centroid
$^{g}$ redshift based on a
single faint line;
$^{\dag}$: sources reported in Fiore et al. 2000}
\end{sidewaystable*}

%% file: h4366.bbl
\begin{thebibliography}{}
\bibitem[]{} Akiyama, M., Ohta, K., Yamada, T., et al. 2000, \apj, 532, 700
\bibitem[]{} Akiyama, M., Ueda, Y., \& Ohta, K. 2002, in the ``AGN
Surveys'', Proceedings of IAU Colloquium 184, editors R.F. Green,
E.Ye. Khachikian, and D.B. Sanders, ASP, p. 245 
\bibitem[]{} Alexander, D.M., Brandt, W.N., Hornschemeier, A.E., et
al. 2001, \aj, 122, 2156
\bibitem[]{} Alexander, D.M., Bauer, F.E., Brandt, W.N., et al. 2003,
AJ, in press (astro--ph/0304392)
\bibitem[]{} Antonucci, R.R.J. 1993, \araa, 31, 473 
\bibitem[]{} Baldi, A., Molendi, S., Comastri, A., et al.
2002, \apj, 564, 190 (Paper I)
\bibitem[]{} Barcons, X., Carrera, F.J., Watson, M.G., et al. 2002, \aap, 382, 522
\bibitem[]{} Barger, A., Cowie, L., Mushotzky, R.F., \&
Richards, E.A. 2001, \aj, 121, 662
\bibitem[]{} Barger A., Cowie L., Brandt, W.N., et al.~2002, \aj, 124, 1839
\bibitem[]{} Bershady, M. A., Lowenthal, J. D., \& Koo, D. C., 1998,
\apj, 505, 50
\bibitem[]{} Bertin, E., \& Arnouts, S. 1996, \aaps, 117, 393
\bibitem[]{} Brotherton, M. S., Tran, H. D., Becker, R. H., et
al.~2001, \apj, 450, 559  
\bibitem[]{} Brandt, W.N., Alexander, D.M., Hornschemeier, A.E., 
et al. 2001 \aj, 122, 2810
\bibitem[]{} Cagnoni, I., Della Ceca, R., \& Maccacaro, T. 1998, 
\apj, 493, 54
\bibitem[]{} Cash, W. 1979, \apj, 228, 939
\bibitem[]{} Ciliegi, P., Vignali, C., Comastri, A., et al.
2003, MNRAS, 342, 575
\bibitem[]{} Comastri, A., Setti, G., Zamorani, G., \& Hasinger, G. 1995, 
\aap, 296, 1   
\bibitem[]{} Comastri, A. 2001, in ``X--ray Astronomy '999:
Stellar Endpoints, AGNs and the Diffuse X--ray Background'', 
ed. N.E. White, G. Malaguti, and G. G.C. Palumbo, AIP conference
proceedings, Vol. 599, p. 73 
\bibitem[]{} Comastri, A., Mignoli, M., Ciliegi, P., et al. 2002a,
\apj, 571, 771 (Paper II) 
\bibitem[]{} Comastri, A., Brusa, M., Ciliegi, P., et al. 2002b, in
``New Visions of the X--ray Universe in the XMM--Newton and Chandra
Era'', ESA SP--488, August 2002 eds. F. Jansen, astro--ph/0203114 
\bibitem[]{} Costamante, L., Ghisellini, G., Giommi, P., et al. 2001,
\aap, 271, 512 
\bibitem[]{625} Daddi E., Cimatti, A., Pozzetti L., et al.~2000, \aap,
361, 535
\bibitem[]{} Dickey, J.M., \& Lockman, F.J. 1990 \araa, 28, 215    
\bibitem[]{} Fabbiano, G., Kim, D.W., \& Trinchieri, G. 1992, \apjs, 80, 531
\bibitem[]{} Fiore, F., La Franca, F., Giommi, P., et al. 2000, NewA, 5, 143 (F00)
\bibitem[]{} Fiore, F., Antonelli, L.A., Ciliegi, P., et al. 2001, in
``X--ray Astronomy '999: Stellar Endpoints, AGNs and the Diffuse
X--ray Background'',  ed. N.E. White, G. Malaguti, and
G. G.C. Palumbo, AIP conference proceedings, Vol. 599, p. 111 
\bibitem[]{} Fiore, F., Giommi, P., Vignali, C. et al. 2001, \mnras, 327, 771
\bibitem[]{} Fiore, F., Brusa, M., Cocchia, F. et al. 2003, \aap, in
press (Paper IV), astro--ph/0306556
\bibitem[]{} The First XMM-Newton Serendipitous Source Catalogue: 1XMM User
Guide to the Catalogue, Release 1.2 15 April 2003, Associated with
Catalogue version 1.0.1, Prepared by the XMM-Newton Survey Science
Centre Consortium (http://xmmssc-www.star.le.ac.uk/) 
\bibitem[]{} Fossati, G., Maraschi, L., Celotti, A., Comastri, A., \&
Ghisellini, G. 1998, \mnras, 299, 433
\bibitem[]{} Freeman, P.E., Kashyap, V., Rosner, R. \& Lamb, D. 2002,
\apjs, 138, 185
\bibitem[]{} Ghisellini, G., Celotti, A., Fossati, G., Maraschi, L., 
\& Comastri, A. 1998, \mnras, 301, 451
\bibitem[]{} Giacconi, R., Rosati, P., Tozzi, P., et al. 2001, \apj, 551, 664  
\bibitem[]{} Giacconi, R., Zirm, A., Wang, J., et al. 2002, \apjs, 139, 369
\bibitem[]{} Gilli, R., Salvati, M., \& Hasinger, G. 2001, \aap, 366, 407
\bibitem[]{} Gunn, K.F., McHardy, I.M., Seymour, N., et al. 2003, 
Astr. Nachr., 324, 105
\bibitem[]{} Hasinger, G., Altieri, B., Arnaud, M., et al. 2001, \aap, 365, L45
\bibitem[]{} Hornschemeier, A.E., Brandt, W.N., Garmire, G.P., et
al. 2000, \apj, 541, 49 
\bibitem[]{} Jarvis, M.J., Rawlings, S., Eales, S., et al. 2001, \mnras,
326, 1585
\bibitem[]{} Kim, D. W.,\& Elvis, M. 1999, \apj, 516, 9 
\bibitem[]{} Koo, D.C., \& Kron R.G. 1988, ApJ, 325, 92
\bibitem[]{} Lehmann, I., Hasinger, G., Schmidt, M., et al. 2001,
\aap, 371, 833 
\bibitem[]{} Lumb, D.H., Guainazzi, M., \& Gondoin, P. 2001, \aap,
376, 387 (L01)  
\bibitem[]{} Madau, P., Ghisellini, G., \& Fabian, A.C. 1994, \mnras,
270, 17
\bibitem{} Mainieri, V., Bergeron, J., Hasinger, G., et al. 2002, \aap, 393, 425 
\bibitem[]{} Maiolino, R., Marconi, A., Salvati, M., et al.2001, 
\aap, 365, 28                   
\bibitem[]{} McCarthy, P.J. 1993, \araa, 31, 639
\bibitem[]{} Mignoli, M., Zamorani, G., \& Marano, B. 2002, in
``Lighthouses of the Universe'', MPA/ESO, p. 590
\bibitem[]{} Mignoli, M.,  et al. 2003, in preparation (Paper V)
\bibitem[]{} Monet, D., Bird A., Canzian, B., et al. 1998, The PMM USNO--A2.0 Catalog, 
(Washington D.C: U.S. Naval Observatory)
\bibitem[]{} Moran, E.C.,  Filippenko, A. V., \& Chornock R. 2002, \apj, 579, L71
\bibitem[]{} Moorwood, A.F.M., Cuby, J.-G., Ballester, P., et
al. 1999, The Messenger, 95, 1
\bibitem[]{} Mushotzky, R.F., Cowie, L.L., Barger, A.J., \& Arnaud,
K.A. 2000, \nat, 404, 459  
\bibitem[]{} Norman, C., Hasinger, G., Giacconi, R., et al. 2002, \apj, 571, 218
\bibitem[]{} Page, M.J., McHardy, I.M., Gunn, K.F., et al. 2003,
Astr. Nachr., 324, 101 
\bibitem[]{} Patat, F. 1999, Efosc2 Users' Manual,
LSO--MAN--ESO--36100--0004 
\bibitem[]{} Piconcelli, E., Cappi, M., Bassani, L., et al. 2002, \aap, 394, 835
\bibitem[]{} Pozzetti, L. \& Madau, P. 2000, in ``The Extragalactic
Infrared Background and its Cosmological Implications'', IAU Symposium
204, p. 71, eds. M. Harwitt \& M. G. Hauser, astro--ph/0011359 
\bibitem[]{} Roche, N., Dunlop, J., \& Almaini, O. 2003, submitted to
MNRAS, astro--ph/0303206 
\bibitem[]{} Saracco, P., Giallongo, E., Cristiani, S., et al.
2001, A\&A, 375, 1
\bibitem[]{} Setti, G., \& Woltjer, L. 1989, \aap, 224, L21
\bibitem[]{} Severgnini, P., Caccianiga, A., Braito, V., et al. 2003,
A\&A, in press, astro--ph/0304308 
\bibitem[]{} Smail, I., Owen, F.N., Morrison, G.E. et al. 2002, \apj, 581, 844
\bibitem[]{} Stern, D., Moran, E.C., Coil, A.L., et al. 2002, \apj, 568, 71
\bibitem[]{} Vignali, C., Mignoli, M., Comastri, A., Maiolino, R.,
\& Fiore, F. 2000, \mnras, 314, L11 
\bibitem[]{} Ueda, Y., Takahashi, T., Ishisaki, Y., \& Ohashi, T. 
1999, \apj, 524, L11
\bibitem[]{} Wilkes, B.J., Schmidt, G.D., Cutri, R.M., et al. 2002, 
\apj, 564, L65
\bibitem[]{} Willott, C., Rawlings, S., Jarvis, M.J., Blundell,
K.M. 2003, MNRAS, 339, 397 
\end{thebibliography}
